\documentclass[12pt,preprint]{aastex}
\usepackage{amsmath}

\shorttitle{DCBs \& GW sources}
\shortauthors{Ablimit \& Maeda}

\begin{document}


\title{Monte Carlo population synthesis on massive star binaries:
Astrophysical implications for gravitational wave sources}
\author{Iminhaji Ablimit\altaffilmark{1} and
Keiichi Maeda\altaffilmark{1} }
\altaffiltext{1}{Department of Astronomy, Kyoto University, Kitashirakawa-Oiwake-cho, Sakyo-ku, Kyoto 606-8502; iminhaji@kusastro.kyoto-u.ac.jp
}


\begin{abstract}

There are important but unresolved processes in the standard formation scenarios of double compact star binaries (DCBs; BH-BH, BH-NS, NS-NS systems), such as mass transfer and the common envelope (CE) phase.
We analyze the effects of different assumptions on key physical processes and binary initial conditions on massive star binary evolution with binary population synthesis (BPS), including a survey of proposed prescriptions for the mass transfer ($\rm q_{\rm cr}$) and the binding energy parameter ($\lambda$) in the CE phase.
We find that $\rm q_{\rm cr}$ clearly affects the properties of NS-NS systems while $\lambda$ has influence on the mass distributions of BH-BH systems. The merger rates of DCBs are increased by efficient CE ejection, which in our prescription is related to the binding energy parameter including all the possible budgets to the energy content. It has been suggested that the difference in the properties of GW150914 and GW151226 may reflect different metallicity. We reproduce their properties with our BPS calculations and find that the property of BH-BH systems at low metallicity is sensitive to $\lambda$; the efficient CE ejection leads to a top-heavy mass distribution both for the primary and secondary BHs, which is favored to explain the nature of GW150914. The efficient CE ejection also leads to enhancement of both the BH-BH and NS-NS merger rates to the level consistent with the observational constraints from the detected gravitational wave sources including GW170817.
\end{abstract}


\section{Introduction}

Double compact star binaries (DCBs), such as Black hole-Black hole (BH-BH), BH-Neutron
star (BH-NS) and NS-NS binaries, play an important role in binary population
synthesis (BPS) studies. They provide excellent laboratories to test physical parameters in binary formation and evolution (Dominik et al. 2012). They are also potential progenitors of various astrophysical
objects, such as ultra-compact X-ray binaries (van der Sluys et al. 2005) and short
$\gamma$-ray bursts (Eichler et al. 1989; Narayan et al. 1992). Furthermore, the inspiral and merger of DCBs have been promising targets as Gravitational wave (GW) emitters within reach by modern GW detectors (Advanced LIGO, Advanced Virgo and KAGRA) (Kulkarni
et al. 1993; Morscher et al. 2013; Tanikawa 2013; Postnov \& Yungelson 2014; Belczynski et
al. 2015; O'Leary et al. 2016). In the year 2015,
gravitational wave sources GW150914 and GW151226 were finally detected by Advanced
LIGO (Abbott et al. 2016a,e). More recently, another massive binary black hole merger GW170104 is also reported by Advanced LIGO (Abbott et al. 2017a). On August 14, 2017, the Advanced Virgo detector and the two Advanced LIGO
detectors coherently observed GWs from the coalescence of two
black holes, GW170814 (Abbott et al. 2017b). These detections gave a crucial proof for Albert Einstein's theory of general relativity (Einstein 1918). Somewhat surprisingly (but see Belczynski et al. 2002), these sources are not the expected NS-NS mergers, but the BH-BH mergers. These detections confirm the existence of BHs, and challenge our understanding of the stellar evolution. Any stellar evolution model must account for the existence of the merging BH-BH binaries within the Hubble time. Observations of NS-NS systems such as the binary pulsar PSR
1913+16 demonstrate that the merging time can be shorter than the Hubble time, making NS-NS mergers promising detectable GW emitters (Phinney 1991).  Eventually, the first GW detection from a NS-NS merger, GW170817, was announced on 17 August, 2017, during the second run (O2) of the LIGO and Virgo Interferometer  (Abbott et al 2017c). These detected GW sources opened a new window for studying DCBs.

Following the discovery of the BH-BH mergers, subsequent works revisited the merger rate of BH-BH binaries
(Abbott et al. 2016c) and studied possible evolution scenarios especially for GW150914.
Belczynski et al. (2016) proposed the standard (isolated)
binary evolution scenario including the common envelope (CE) phase at a low metallicity for GW150914.
Woosley (2016) discussed possible evolutionary scenarios for GW150914, considering both a rapidly-rotating single star model and a binary model. He claimed that the characteristics of GW150914 are unlikely to be explained by any single-star model, and the binary model is more favored.
On the other hand, Tagawa et
al. (2016) presented post-Newtonian N-body simulations of mergers of gas-accreting stellar-mass
black holes (BHs) within dense stellar environments, and claimed that the BH-BH
merger GW150914 was likely driven by three-body encounters accompanied by a few $M_\odot$ of
gas accretion.
GW151226 differs from GW150914 in the significantly
lower BH masses (Abbott et al. 2016e). BHs with such masses can be formed at solar
metallicity (Spera et al. 2015). As we see from these possibilities, we still do not
know exact evolutionary path ways to the
compact star mergers (e.g., the merging BHs like GW150914).

In studying the natures of DCBs, obtaining reliable predictions of their merger rates
and birth rates is still a challenge (Abadie et al. 2010). For example, the mass transfer
(and/or mass loss) process, the CE phase and some other parameters
in the binary evolution are critically important but the details are not well known. These
can be tested by accumulating samples of GW detections from BH-BH and NS-NS systems as well as expected detections of BH-NS systems. Advanced LIGO, Advanced Virgo and
KAGRA (Harry et al. 2010; Sengupta et al. 2010; Somiya 2012) will probe the universe in
search for DCB signatures. To be ready for such detections from a theoretical point view, we
still need to continue to present more information on the formation and evolution of DCBs.
In the past decades, a series of Binary Population Synthesis (BPS) works on DCBs have
been performed to investigate the effects of various physical parameters on the formation,
evolution and merger rate of DCBs, but there are still debates on the treatment of uncertain
physical processes.

The mass transfer process and the CE phase are crucial for producing all kinds of compact star binaries, while the details of these processes are still unknown. Massive MS/MS close binaries (where MS means a main-sequence
star, with an initial mass $\gtrsim 8M_\odot$) generate close compact star binaries mainly by going through a CE phase
at least once. The MS/MS binary may involve the CE phase if the initial orbital separation is not large enough and the more massive one fills the Roche lobe (RLOF) due to the expansion of the star during rapid nuclear evolution. If the mass transfer time scale is shorter than the thermal time scale of the accretor, the CE is a likely outcome where the extended envelope engulfs both stars. Especially, the first mass transfer may occur on the dynamical timescale of the primary star for close systems, and then it will lead to unstable mass transfer and the CE evolution. If the stars survive the CE phase, then the binary may again evolve into CE evolution including a BH/NS, if the second RLOF mass transfer proceeds on the dynamical timescale. The system that survived the CE phase would finally become a DCB. The MS/MS binaries in a wider orbit may also evolve into CE evolution after the massive one evolved to a BH/NS if the second RLOF mass transfer is unstable, and the DCBs would be produced after the successful CE ejection (Iben \& Livio 1993; Ivanova et al.2013). As the orbital energy and angular momentum are removed by the CE ejection (Paczy\'{n}ski 1976), the binary separation is decreased after the CE phase.
Details of the CE phase have however not been well understood due to its short life time, despite some works which address indirect observational
evidence for the CE evolution (e.g., Sion et al. 2012; Ivanova et al. 2013).

Theoretical studies still contain uncertainties. However, the theoretical predictions are crucial to understand astrophysical natures of
the progenitor systems of detected GW sources (Stevenson et al. 2015). The related
estimates and studies have been presented by different groups in the past decade, with
different codes and different sets of physical assumptions (e.g. Lipunov et al. 1997; Nelemans et al.
2001; Voss \& Tauris 2003; Dewi \& Pols 2003; Yungelson et al. 2006; Belczynski et al. 2008;
Mennekens \& Vanbeveren 2014; Eldridge \& Stanway 2016). The difficulty of reconciling the disparate results, and many different assumptions made in these works, highlight the uncertainty in modelling key stages of stellar evolution. It is thus always useful to
add and compare the results of different approaches to the problem. de Mink \&
Belczynski (2015) performed a comparative BPS study on the evolutionary path of massive
stars by adopting a similar approach as Dominik et al. (2012), varying some important initial
inputs (such as initial mass function, distribution of mass ratio, distribution of orbital periods) based on observation results of Sana et al. (2012). They further provide the merger rates and distributions of properties for BH-BH, BH-NS and NS-NS systems by adopting the binding energy parameter from Xu \& Li (2010) for the CE and assuming that half of the mass is accreted but the other half is ejected from the system in their mass transfer prescription. In this paper, we further expand the exploration on how different treatment of these processes would affect the outcome.
We use three proposed expressions for the binding energy parameter in the CE model ($\lambda_{\rm g}$, $\lambda_{\rm b}$, and $\lambda_{\rm e}$, the values derived from the MESA, see \S 2 for details, and Wang et al. 2016a,b; Ablimit et al. 2016), and adopt three detailed mass transfer models (three different $q_{\rm cr}$, for more details see Section 2).

In this paper, we investigate influences that a series of binary evolution prescriptions
($q_{\rm cr}$, $\lambda$, $\alpha$, eccentricity, initial separation, kick velocity
distribution and other parameters) have on the formation of DCBs (see Table 1).
We present the properties of the possible gravitational wave signals from the mergers of
those DCBs in our BPS study. In \S 2, we describe our BPS code and our models to treat
the binary physical processes including the CE evolution and mass transfer assumptions.
The properties of DCBs and merging systems produced by our models are presented in \S 3. In \S 3, we also present implications for the nature of the detected GW sources, including GW150914 and GW151226 (BH-BH), and GW170817 (NS-NS).  
The paper is closed in \S 4 with conclusions and discussion.

\section{Monte Carlo binary population synthesis}
\label{sec:model}

Our BPS code is the one developed by Hurley et al. (2002) and modified by Kiel \& Hurley (2006). As compared to the version used by Kiel \& Hurley (2006), we have updated and modified the code in several aspects, especially regarding the conditions for mass transfer and the treatments of CE evolution, which are briefly described below. For the initial conditions, we consider the constraints from previous observational results, including new constraints for massive O stars by Sana et al.(2012) (i.e., high binary fraction, preferentially short orbital periods and uniform mass ratio distribution). We adopt the initial mass function (IMF) of Kroupa et al. (1993) for the primary mass distribution,
\begin{equation}
f(M_1) = \left\{ \begin{array}{ll}
0 & \textrm{${M_1/M_\odot} < 0.1$}\\
0.29056{(M_1/M_\odot)}^{-1.3} & \textrm{$0.1\leq {M_1/M_\odot} < 0.5$}\\
0.1557{(M_1/M_\odot)}^{-2.2} & \textrm{$0.5\leq {M_1/M_\odot} < 1.0$}\\
0.1557{(M_1/M_\odot)}^{-\alpha} & \textrm{$1.0\leq {M_1/M_\odot} \leq 150$},
\end{array} \right.
\end{equation}
with $\alpha = 2.7$ in this study. The secondary mass distribution is determined by the distribution of the initial mass ratio,
\begin{equation}
n(q) = \left\{ \begin{array}{ll}
0 & \textrm{$q>1$}\\
\mu q^{\nu} & \textrm{$0\leq q < 1$},
\end{array} \right.
\end{equation}
where $q=M_2/M_1$, $\mu$ is the normalization factor for the assumed power law distribution with the index $\nu$. We consider a flat distribution ($\nu = 0$ and $n(q)=$constant) for the initial mass ratio distribution (IMRD). The distribution of the initial orbital separation, $a_{\rm i}$, is assumed to be given by the following formalism
(Davis et al. 2008),
\begin{equation}
n(a_{\rm i}) = \left\{ \begin{array}{ll}
0 & \textrm{$a_{\rm i}/R_\odot < 3$ or $a_{\rm i}/R_\odot > 10^{6}$}\\
0.078636{(a_{\rm i}/R_\odot)}^{-1} & \textrm{$3\leq a_{\rm i}/R_\odot \leq 10^{6}$} \ .
\end{array} \right.
\end{equation}

We test two possibilities for the orbit eccentricity. In one case, the uniform (flat) initial eccentricity distribution is assumed in a range between 0 to 1.
We also test the possibility that binaries are in circle orbits or have negligible eccentricities (Dominik et al. 2015; Nishizawa et al. 2016) with tighter orbital distribution $3\leq a_{\rm i}/R_\odot \leq 10^4$ (Hurley et al. 2002). This is motivated by observational results of Sana et al.(2012). We assume the binary fraction of $100\%$ (${f_{\rm bin}} = 1$). We fix the metallicity to be $Z=0.02$ for the parameter study for the key binary evolution processes, while we consider the low metallicity case of $Z=0.001$ in discussing the origins of GW150914 and GW151226 (\S 3.3).

Regarding the key physical processes, our simulations have three main tunable parameters, i.e.,
$\alpha_{\rm CE}$, $\lambda$ and $q_{\rm cr}$. One main aim of this paper is to investigate the effects that these prescriptions have on the evolution toward compact star binaries.
The CE evolution is an important but unsolved phase in the binary evolution process.
The widely used $\alpha$-model considers energy conservation (Webbink 1984),
\begin{equation}
E_{\rm{bind}} = {\alpha_{\rm CE}} \Delta E_{\rm orb} \ ,
\end{equation}
where $E_{\rm{bind}}$, $\alpha_{\rm CE}$, and $\Delta E_{\rm orb}$ are the binding energy of the envelope,
the efficiency parameter and the change in
the orbital energy during the CE phase, respectively.

In the $\alpha$-formalism, the efficiency parameter ($\alpha_{\rm CE}$) and the binding energy parameter ($\lambda$) are vital to determine the fate of the CE evolution, but they have been treated as constants in most previous works. However, in reality, they should change with the mass and evolutionary stage of the primary star.
The binding energy of the envelope
is expressed by the following:
\begin{equation}
E_{\rm{bind}} = - \frac{GM_1 M_{\rm{en}}}{\lambda {R}_1},
\end{equation}
where $M_1$, $M_{\rm{en}}$ and ${R}_1$ are the total mass, envelope
mass and radius of the primary star, respectively. The two stars coalesce if the radius of either of the stars satisfies Equation (75) of Hurley et al.(2002) and if the CE phase is longer than the dynamical timescale (for details of the criterion for surviving or merging during the CE phase, see Hurley et al.2002).

The value of $\lambda$ for evolving stars can be calculated by considering gravitational energy only (hereafter $\lambda_{\rm g}$), adding inner energy ($\lambda_{\rm b}$), or adding entropy of the envelope ($\lambda_{\rm e}$).
The binding energy can be described as follows if it is only due to the gravitational binding energy,
\begin{equation}
E_{\rm{bind}} = \int^{M_1}_{M_c}{-\frac{GM(r)}{r}}\mathrm{d}m \ ,
\end{equation}
where $G$ is the gravitational constant, $M_1$ is the primary mass and $M_c$ is its
core mass. If both of the thermal energy and the recombination energy contribute to the binding
energy, then
\begin{equation}
E_{\rm{bind}} = \int^{M_1}_{M_c}{[-\frac{GM(r)}{r} + U]}\mathrm{d}m \ ,
\end{equation}
where $U$ is the internal energy (Han et al. 1994). If the enthalpy model (Ivanova \& Chaichenets
2011) is used, then
\begin{equation}
E_{\rm{bind}} = -\int^{M_1}_{M_c}{[-\frac{GM(r)}{r} + U + \frac{P}{\rho}]}\mathrm{d}m \ ,
\end{equation}
where $P$ and $\rho$ are the pressure and the density of the gas, respectively. For more details, see Ivanova \& Chaichenets (2011) and Wang et al. (2016a,b). We note that the structure parameters need to be calculated by a detailed stellar evolution code (e.g. Eldridge et al 2017). We test all the three prescriptions independently in our BPS study, adopting the calculation results with the MESA code by Wang et al. (2016a,b).  Wang et al. (2016b) discussed that the MESA code is more powerful in probing the stellar structure than Eggleton (1971)'s stellar evolution code previously adopted by Xu \& Li (2010).  The calculated values of binding energy parameters by Wang et al. (2016b) are remarkably different from those of Xu \& Li (2010) in specific circumstances. They concluded that the $\lambda$-value varies when a star evolves and strongly depends on the star's initial mass, and $\lambda_{\rm e}$ is several times larger than $\lambda_{\rm b}$ and $\lambda_{\rm g}$, which can assist the ejection of the CE. This is a new and more physically motivated treatment, comparing to previous BPS studies (while all previous works treated $\lambda$ by adopting Xu \& Li (2010) or similar works, or even took it as a constant).

Most BPS studies take $\alpha_{\rm CE}$ merely as a constant (typically 1 or 3, see Hurley et al. 2002 and Dominik et al. 2012). For relatively low mass stars, Davis et al. (2012) derived a relation of $\alpha_{\rm CE}$ with $q$ from analysis of observed post-CE white dwarf (WD) binaries. On the other hand, a relation between $\alpha_{\rm CE}$ and $q$ for massive stars leading to a NS/BH DCBs is not clarified yet. As a demonstration, in this paper we test the description by Davis et al. (2012) for $\alpha_{\rm CE}$. While it would not provide a physically motivated relation for the massive stars, it would give an estimate about how the results are dependent on variations of $\alpha_{\rm CE}$. In typical situation, the value of $\alpha_{\rm CE}$ calculated from this relation is lower than 1 (see Davis et al. 2012 and Ablimit et al. 2016), similar to a constant value adopted for massive binary evolution. We adopt the following formula from Davis et al. (2012):
\begin{equation}
{\rm log_{10}}{\alpha_{\rm CE}} = {\epsilon}_0 + {{\epsilon}_1}{\rm log_{10}}{(q)} \ ,
\end{equation}
where ${\epsilon}_0$ and ${\epsilon}_1$ are constants taken from Davis et al. (2012). We simulate our BPS by treating the CE phase with this equation or with ${\alpha_{\rm CE}=1}$ for the comparison with other theoretical assumptions.

The critical mass ratio ($q_{\rm cr}$) is another key physical parameter that determines the stability of the mass transfer. Shao \& Li (2012) computed the critical mass ratio considering the
possible response of the accreting star (i.e., spin-up and rejuvenation) under three different assumptions (including conservative and non-conservative cases): (1) Half
of the transferred mass is accreted by the secondary, and the
other half is lost from the system, also removing the specific orbital
angular momentum of the accretor (see also de Mink et al. 2007). (2) The transferred mass is assumed to be accreted
by the secondary unless its thermal timescale (${\tau_{\rm KH_2}}$) is much
shorter than the mass transfer timescale ($\tau_{\dot{M}}$). The accretion
rate is limited by--$[\rm{min}(10(\tau_{\dot{M}} /{\tau_{\rm KH_2}}), 1)]\dot{M}_1$
(Hurley et al. 2002). Rapid mass accretion may drive the accretor
out of thermal equilibrium. It will then expand and become
over-luminous. Shao \& Li (2012) found the values of ${\tau_{\rm KH_2}}$ are usually much lower
than that of the same star in thermal
equilibrium. Therefore, it is always as ${\tau_{\rm KH_2}} < 10{\tau_{\dot{M}}}$, and the
mass transfer is generally conservative. (3) The accretion
rate onto a rotating star is reduced by a factor of ($1 - {\Omega}/{\Omega_{\rm cr}}$),
where $\Omega$ is the angular velocity of the star and ${\Omega_{\rm cr}}$ is its critical
value. In this prescription, a star cannot accrete
mass when it rotates at ${\Omega_{\rm cr}}$. The remaining material is ejected out of
the binary by the wind, and it takes away the
specific orbital angular momentum from the accretor. The critical mass ratios corresponding to these three cases are
denoted as $q_{\rm cr1}$, $q_{\rm cr2}$ and $q_{\rm cr3}$, respectively.
For each model as described above, we run the BPS simulations with these different treatment of the
critical mass ratio (i.e., adopting $q_{\rm cr} = q_{\rm cr1}$, $q_{\rm cr2}$, or $q_{\rm cr3}$). The three mass transfer models also represent a major update in our BPS code.

For $q_{\rm cr3}$, we track $\Omega$ (the angular rotational velocity of a star) in the same way as given in Hurley et al. (2000). We adopt a uniform distribution of the rotational velocity within a star (van den Heuvel 1968). The spin evolution of each component in a close binary system could be influenced by tidal synchronization, which is taken into account in our calculations. For the evolution of angular velocity of each star, including the tidal evolution of a binary, we follow default prescriptions in the BPS code of Hurley et al.(2002).

Another main issue is the evolution of `single' massive stars and the birth of the compact stars. For massive stars, stellar winds have non-negligible
influence on final outcomes (Belczynski et al. 2010), but their loss rates have not been accurately determined. For example, Hurley et al. (2000) (wind 1) and Vink et al. (2001) (wind 2) gave different prescriptions for O and B stars in different stages. Here, we adopt two prescriptions for the wind mass loss rate: the wind 1 model based on Hurley et al. (2000,2002) and wind 2 model based on Vink et al. (2001) and Vink \& de Koter (2002). For low mass H-rich stars, it is taken from Hurley et al. (2000). Another main difference between the  default model wind 1 and the wind 2 model exists for treatment of luminous blue variable stars (LBVs). In the wind 1 for LBVs with $L/L_\odot > {6\times10^5}$ and ${10^{-5}({R/R_\odot})(L/L_\odot)^{0.5}} > 1$:
\begin{equation}
{\dot{M}} = 0.1\times{[10^{-5}({R/R_\odot})({L/L_\odot})^{0.5} - 1]}^3(\frac{(L/L_\odot)}{6\times10^5} - 1)\, M_\odot{\rm{yr}}^{-1} ,
\end{equation}
where $R$ and $L$ are the radius and luminosity of the star, respectively (see Humphreys \& Davidson 1994, Hurley et al. 2000, for more details). This equation applies to the time-average mass loss rate from LBVs including the eruptive phase. A simpler prescription is adopted in the wind model 2,
\begin{equation}
{\dot{M}} =  f_{lbv}\times10^{-4}\, M_\odot{\rm{yr}}^{-1} ,
\end{equation}
with the standard factor $f_{lbv} = 1.5$ (also see Vink \& de Koter 2002). This value is also within the (uncertain) range of the time-average LBV mass loss rate taking into account the eruptive phase. The mass loss rate of LVBs is highly uncertain. The observed LBV-mass loss rates are large, ranging between $10^{-5}$ and $10^{-3} M_\odot$ yr$^{-1}$ (Humphreys \& Davidson 1994; Vink \& de Koter 2002; Davidson \&Humphreys 2012). It has been suggested that the effect of the eruptive mass loss is important in the formation of DCBs  (Mennekens \& Vanbeveren 2014). To further test this effect, we also run the BPS simulation for a specific model with the LBV mass loss rate in equation 11 replaced by 0 (i.e., the LBV wind is switched off) or $10^{-3} M_\odot{\rm{yr}}^{-1}$, which will be discussed in \S 3.1.

Supernova (SN) explosions play an important role on the birth of BHs and NSs. We consider the prescription which successfully reproduces the observed mass distribution of BHs and NSs. This is based on the neutrino-driven
convection-enhanced SN mechanism (the so-called rapid SN
mechanism). This prescription results in a `failed' SN for a massive star with the initial mass $> 20-25 M_\odot$ giving the birth of a BH. The gravitational mass of the compact remnant under the rapid SN mechanism is described by Fryer et al. (2012) as follows,
\begin{equation}
M_{\rm rem} = 0.9(M_{\rm proto} + M_{\rm fb}) \ .
\end{equation}
Here $M_{\rm proto} = 1.0M_\odot$ is the mass of the proto-compact object and
\begin{equation}
M_{\rm fb} = \left\{ \begin{array}{ll}
0.2M_\odot & \textrm{${M_{\rm CO} } < 2.5M_\odot$}\\
0.286{M_{\rm CO}} - 0.514M_\odot & \textrm{$2.5M_\odot < {M_{\rm CO} } < 6M_\odot$}\\
f_{\rm fb}(M_1 - M_{\rm proto}) & \textrm{${M_{\rm CO}} \geq 6M_\odot$},\\
\end{array} \right.
\end{equation}
with
\begin{equation}
f_{\rm fb} = \left\{ \begin{array}{ll}
1.0 & \textrm{$6M_\odot \leq {M_{\rm CO} } < 7M_\odot$}\\
{a_{1}}({M_{\rm CO} / M_\odot}) + b_1 & \textrm{$7M_\odot \leq {M_{\rm CO} } < 11M_\odot$}\\
1.0 & \textrm{${M_{\rm CO}} \geq 11M_\odot$},\\
\end{array} \right.
\end{equation}
where $M_{\rm CO}$ is the CO core mass, $M_1$ is the mass of the progenitor star mass just before the core-collapse, $a_{1} = 0.25 - (1.275{M_\odot}/(M_1 - M_{\rm proto}))$ (see Wang et al. 2016a) and $b_1 = -11a_{1} + 1$.

The formation of a BH is connected to the compactness of the stellar core at the
time of collapse in this prescription. Recent study on the SN explosion mechanism suggests that the compactness parameter is an important parameter, which is defined as a mass within a given radius in the core of the progenitor star (O'Connor \& Ott 2011). A star with a small compactness parameter is more likely to explode
as an SN and produce a NS, while a star with a large compactness parameter tends to evolve to the failed SN that produces a BH. The He core mass or CO core mass prior to core collapse determines the compactness parameter, thus the newly born BH mass (e.g. Kochanek 2014, 2015; Sukhbold \& Woosley 2014; Clausen et al. 2015).
In our BPS simulation, the He and CO core masses are tracked following Hurley et al. (2000). We require that $M_{\rm BH} \geq 3M_\odot$.  We do not consider the formation of BHs by an accretion-induced collapse (AIC) of a NS in X-Ray Binaries. On the other hand, the possibility of the NS formation through the electron-capture SN is taken into account in our recipe (Fryer et al. 2012), and the NS formation by the AIC of a WD is also taken into account in our calculations (Nomoto \& Kondo 1991; Ablimit \& Li 2015). The default values of Hurley et al.(2002) are used for the maximum masses of the NS and WD.

We also assume that a natal kick is imparted on the newly born BHs, similar to the case of the NS formation. The kick velocity is set to be inversely proportional to the remnant mass, i.e., ${\upsilon}_{\rm k}(\rm BH) = (3M_{\sun}/M_{\rm BH}){\upsilon}_{\rm k}(NS)$, where ${\upsilon}_{\rm k}(\rm NS)$ is the kick
velocity for NSs, which follows the Maxwellian distribution as
\begin{equation}
P({\upsilon}_{\rm k}) = \sqrt{\frac{2}{\pi}}\frac{{{\upsilon}_{\rm k}}^2}{{{\sigma}_{\rm k}}^3}e^{{{-{{\upsilon}_{\rm k}}^2}}/{2{{\sigma}_{\rm k}}^2}} \ .
\end{equation}
We use the velocity dispersion of ${\sigma}_{\rm k} = 190\,{\rm km\,s^{-1}}$ (Hansen \& Phinney 1997)
or  ${\sigma}_{\rm k} = 265\,{\rm km\,s^{-1}}$ (Hobbs et al. 2005).

In Table 1, we summarize our models, where the `case' distinguishes different combinations of $\lambda$ and $q_{\rm cr}$, while `model' variants are for the other parameters (see Table 1).
In our standard model, we assume ${\alpha_{\rm CE}=1}$, ${\lambda=0.5}$ for the CE phase, $Z=0.02$, a flat IMRD for the initial conditions. For $q_{\rm cr}$ we adopt the default prescription in the BSE code by Hurley et al. (2002) in the standard run. In our standard run, the Belczynski et al.(2002)'s prescription for the SN mechanism is used. For the other model parameters in the standard run, we adopt the values used in model 3. Note that the tidal evolution is taken into account in all the simulations. Combining models 1--3 with cases 1--9, there are thus 27 different models. Including models 4--6 with cases 1--3 and the standard model, we cover 37 different models in our numerical calculations. The default values from Hurley et al.(2002) are adopted for other physical parameters which are not mentioned in this paper.

We calculate our merger rates as follows. We assume a constant
star formation rate (SFR) of $3.5\,{M_\odot}{\rm{yr}^{-1}}$ over the past 10 Gyr. This is a simplified but frequently adopted treatment in the field of BPS study given that the star formation history of our galaxy is not well known (Wyse 2009). To allow straightforward comparisons to previous studies, we have decided to adopt this simple model for the SFR (see Dominik et al. 2012). 

\section{Results }

We perform the binary population synthesis simulations to see the effect of different combinations of the CE parameters,
the critical mass ratio models and other parameters, on the property distributions of the
BH-BH, BH-NS and NS-NS systems and their mergers. We use the BPS code to generate an initial set of ${10}^7$ MS/MS binaries under different models with different cases as described in Table 1 for obtaining these DCBs. 
The initial properties of the $10^7$ individual systems are stochastically constructed in a continuous parameter space following the underlying IMF, period and mass ratio distributions. The primary mass distribution follows the adopted Krupa IMF. The secondary mass is chosen from the mass ratio distribution, independently from the IMF. For each set of the initial parameters, we evolve the binary system to an age of the Hubble time, or until it is destroyed. The metallicity is fixed to be $Z=0.02$ in \S 3.1-3.2, while a low metallicity is also considered in \S 3.3-3.4. Our Monte Carlo simulations have statistical fluctuations due to their finite size. The relative statistical error (${\sqrt{N}}/N$, where $N$ is the number of simulated DCBs who can merge in 10 Gyr) is estimated to be $\sim$ 5\% - 35\% based on the results of model 1--6. The range of the statistical fluctuation here represents the range of frequencies of the systems under consideration; the BH-BH or BH-NS merger as rare events suffer from the largest Monte Carlo noise (up to $\sim 35$\%), while the Monte Carlo noise is much smaller for the NS-NS mergers ($\sim 5$\%).

\subsection{Merger rates}

In Table 2 we list the galactic merger rates of DCBs calculated for the fiducial Milky Way-like galaxy for a single metallicity ($Z=Z_\odot$) and different variations of models. This is visualized in Figure 1 (which also includes the low metallicity cases to be discussed in \S 3.3). The merger rates of BH-BH, BH-NS and NS-NS systems in our standard run are 15.1 ${\rm Myr}^{-1}$, 1.21 ${\rm Myr}^{-1}$ and 40.3 ${\rm Myr}^{-1}$, respectively.

From Figure 1, we observe the general trends with which the different prescriptions for $\lambda$ (binding energy) and $q_{\rm cr}$ (mass transfer) affect the resulting merger rates. First, the variation in the merger rates depending on the different prescriptions is about a factor of a few; therefore it would not drastically change qualitative behavior, but these effects should be taken into consideration if one tries to present quantitative analysis. Second, we have obtained an insight into how these processes affect the resulting merger rates of the BH-BH and NS-NS systems. Adopting $\lambda_{\rm e}$ leads to more efficient CE ejection, therefore resulting in larger DCB formation rates and larger merging rates, both for the BH-BH and NS-NS systems. The BH-BH system is less strongly affected by this prescription, since the binding energy of the CE here is large relative to the small change introduced by different prescriptions in $\lambda$. The prescription for $q_{\rm cr}$ is important to determine the relative ratio of the BH-BH and NS-NS systems. Generally, the mass transfer is less efficient in $q_{\rm cr1}$ and $q_{\rm cr3}$ than $q_{\rm cr2}$ (conservative case). Therefore, the number of the DCB systems leading to the merger is increased with these prescriptions. Especially, with the prescription $q_{\rm cr3}$, in which the spin-up of the accretor is assumed to reduce the net accretion rate,  the binary NS formation is enhanced. 
With our model 4/case 1, we have the highest merger rate of BH-BH systems (27.5 ${\rm Myr}^{-1}$), by the combination of the tighter initial separation, eccentricity with 0, varying $\alpha_{\rm CE}$, $\lambda = \lambda_{\rm e}$ and $q_{\rm cr1}$. 
The highest rates of BH-NS and NS-NS systems (13.8 ${\rm Myr}^{-1}$ and 153 ${\rm Myr}^{-1}$) are produced at $Z_\odot$ by model 4 with constant $\alpha_{\rm CE}$, $\lambda = \lambda_{\rm e}$ and $q_{\rm cr3}$. Namely, the prescriptions of $\alpha_{\rm CE}$ and $\lambda$ affect the rates of all the DCB merger rates in a similar way, and the treatment of $q_{\rm cr}$ affects the merger rates of the BH-BH and the other systems (BH-NS and NS-NS) in the opposite way.

Our wind models take into account the LBV eruptive mass loss phase, and we test different mass loss rate for the LBV phase such as 0, $1.5\times10^{-4}$ and $10^{-3}$ ${\rm Myr}^{-1}$ under the same model. There is no clear difference found in the resulting merger rates between the three prescriptions (see Figure 1). This result is actually similar to the result of Mennekens \& Vanbeveren (2014) under the SN mechanism of Fryer et al. (2012). Thus, we conclude that the LBV wind mass loss rate is not so important for the rate of DCBs. The treatment of the LBV mass loss rate, however, may affect the final natures of BH-BH binaries as a function of the mass, while the effect on the rate is relatively minor. This issue will be addressed further in \S 3.2.

As shown in the dotted region of Figure 1, we obtain the binary BH merger rates of
$\sim 40-63$ ${\rm Myr}^{-1}$ for models 4--6 with $Z=0.001$, for $\lambda = \lambda_{\rm e}$ and $\lambda_{\rm g}$. A similar rate is also predicted for the `standard' run with $Z=0.001$ (with constant $\lambda$ and $q_{\rm cr}$ given by Hurley et al. 2002). At the low metallicity, the predicted NS-NS merger rates are 150--240 ${\rm Myr}^{-1}$ with $\lambda_{\rm e}$, while this is by up to a factor of two smaller with $\lambda_{\rm g}$ and constant $\lambda$. Combining all results (Pop I and II), the resulting range of the BH-BH merger rate is covered by the previous studies (e.g., Mennekens \& Vanbeveren 2014; de Mink \& Belczynski 2015; Eldridge \& Stanway 2016; Belczynski et al. 2016; Chruslinska et al. 2017), while we are at the high rate similar to Belczynski et al. (2016) and Chruslinska et al. (2017). The NS-NS merger rate in our models is generally higher than the other works (\S 3.4 for details), except for a particular model `C+P' presented by Chruslinska et al. (2017). Interestingly, the NS-NS rate in our models increases for decreasing metallicity (\S 3.4), which is similar to some of the previous works (e.g., Hurley et al. 2002; Eldridge \& Stanway 2016), while this particular model with the high NS-NS merger rate by Chruslinska et al. (2017) has the opposite trend. As for the BH-NS merger rate, Eldridge \& Stanway (2016) predicted a higher rate than the predicted BH-NS merger rates in this BPS study.

It is seen that the CE, wind mass loss, SN mechanism, initial separation and mass transfer are crucial for the massive star binary evolution. Especially, a tight initial separation (orbital period) distribution and $\lambda = \lambda_{\rm e}$ increase the merger rates of all kinds of DCBs, and the mass transfer prescription ($q_{\rm cr}$ ) is crucial to determine the relative rates of BH-BH and NS-NS mergers. 
As for the mass transfer, we note that we are not claiming that the rotation-induced mass-loss is a main agency of the mass-loss process; the strong wind prevents the accretion anyway  (e.g., Petrovic et al. 2005; Cantiello et al. 2007). However, our result suggests that the reduction of the mass transfer rate due to the spin of the accretor, if it would happen (while this is uncertain), is still important to determine the final fate of the binary systems, since it could enhance the binary NS survival/formation. At the same time, the prescription $\lambda = \lambda_{\rm e}$ helps to enhance the envelope ejection and thus increases the chance that DCBs survive the CE. The expected rate of merging DCBs is maximized for the initial orbital separations of $3\leq a_{\rm i}/R_\odot \leq 10^4$ instead of $3\leq a_{\rm i}/R_\odot \leq 10^6$.


The initial eccentricity distribution does not have clear effect on merger rates. If the distribution of orbital angular momentum is used to determine the initial state of each binary, rather than one of semi-major axis or period, the results do not depend on the form of any chosen eccentricity distribution. In fact, eccentricity need not be a free parameter. This becomes evident when considering that tidal interaction conserves angular momentum and that almost all systems circularize before RLOF.

\subsection{\textbf{Properties of the DCBs at the formation} }

Figures 2 shows the orbital period distributions of BH-BH, BH-NS and NS-NS systems (upper, middle and lower panels, respectively).
From the results of models 1, 3 and 6, we see that the orbital period is mostly affected by the kick velocity. The model with $\sigma = 190 \rm{Km/s}$ produces a larger number of very short orbital period DCBs than the model with $\sigma = 265 \rm{Km/s}$. A larger number of BH-NS and NS-NS systems in the short orbits are produced with $\lambda = \lambda_e$ and also with $q_{\rm cr} = q_{\rm cr3}$. From the results,
it is seen that the initial eccentricity distribution has small influence on the orbital period distributions.
The two prescriptions of $\alpha_{\rm CE}$ and other initial conditions do not have a significant effect.

The chirp mass distributions of BH-BH, BH-NS and NS-NS systems are given in Figures 3, 4, and 5, respectively. The chirp mass is given as follows:
\begin{equation}
M_{\rm chirp} = {\mu}^{3/5} M^{2/5} \ ,
\end{equation}
where $M = M_1 + M_2$ and $\mu = M_1 M_2/(M_1 + M_2)$ (with $M_1$ and $M_2$ the masses of the primary and secondary, respectively).
The wind mass loss is the most important function to determine the chirp mass of BH-BH systems. With the wind 1 model, the chirp mass of BH-BH systems
ranges only from 5 $M_\odot$ to 15 $M_\odot$, while it ranges from 4 $M_\odot$ to 20 $M_\odot$ with wind 2 model (see Figure 3). 

The upper limits of the chirp mass in BH-BH systems in our calculation (15 and 20 $M_\odot$ with wind 1 and 2 models) are larger than the results of Belczynski et al.(2002, 2010) ($\sim$ 11 and 15 $M_\odot$ with wind 1 and 2 models). In order to clarify the reason which causes this difference in the chirp mass distribution, we additionally test a set of $10^6$ MS-MS binaries and single stars evolution (SSE). For the SSE simulation, we use the SSE code developed by Hurley et al.(2000). We compare the two prescriptions for the SN mechanism (the fate of the core collapse), one by Fryer et al.(2012) (F12, adopted in this paper) and the other one used by Belczynski et al.(2002, hereafter B02). We adopt $\alpha_{\rm CE} = 1$ and $\lambda = 0.5$ for the CE phase, the wind 2 model and solar metallicity in this exercise to compute the BH mass distributions. The  other initial conditions are same as default ones in the code (for more details see Hurley et al. 2000, 2002). The calculated remnant mass from the single star evolution shows that the star which has a initial mass between 22 and $\sim$ 34 $M_\odot$ could leave a more massive remnant under the rapid SN explosion mechanism of F12 than that of B02 (Figure 6). This mass range dominates the formation of the BH-BH systems following the adopted IMF. The mass distributions of the primary BHs in the BH-BH binaries under two SN mechanisms are given in the upper and middle panels of Figure 6. We find that the rapid SN explosion mechanism of F12 could produce more massive BHs than the results under B02. Eldridge \& Stanway (2016) also showed the difference of the BH masses in their calculation and the previous study by Belczynski et al. (2002, 2010), and claimed that the remnant mass calculation likely contributes to the difference. The consideration here demonstrates that the BH mass distribution is sensitive to the treatment of the SN mechanism (which determines the BH mass as a function of the core mass) and the wind mass loss (which determines the mass of the pre-collapse progenitor star). As such, an increasing sample of the GWs from DCBs merging systems may tell this important information to understand the debated fate of massive stars and the unresolved mechanism of SN explosions.

Recipes for the wind mass loss, SN explosion mechanism, initial circular orbit and eccentric orbit affect the nature of the BH-BH populations. Other parameters have little influence on the chirp mass of BH-BH systems. The chirp mass distribution of BH-NS systems tends to peak in a smaller value if $\lambda$ changes with the stellar evolution rather than it is treated merely as a constant (see Figure 4).
Most of the parameters affect the nature of the BH-NS population, but there is not a clear trend in the dependence of the chirp mass distribution for different parameters. In Figure 5, it is shown that most of the parameters do not affect the chirp mass distribution of the NS-NS systems, except for the prescription of $\alpha$. If one uses equation 9 for $\alpha$, the resulting NS-NS systems tend to be massive. While this is taken merely as indicative as our prescription is not calibrated by massive star binaries, this suggests that the value of $\alpha$ is important in determining the mass distribution of the NS-NS systems.

We concluded that the LBV wind mass loss rate has small effect on the merger rate when the Fryer's rapid SN explosion model is adopted (\S 3.1), However, this does not mean that the LBV wind mass loss is unimportant to shape the nature of the DCBs. Figure 7 shows the chirp mass distributions obtained with different LBV wind mass loss rates. It is seen that the wind mass loss of LBVs is indeed important to determine the masses of the BH-BH systems.

\subsection{Behaviors at low metallicity: Implications for GW150914-like and GW151226-like binary black hole mergers}

GW150914 has been identified as a merging BH-BH binary.
The estimated pre-merging primary and secondary BH masses are in the
ranges of $31$ -- 41 $M_\odot$ and $25$ -- 33 $M_\odot$ (with a chirp mass around $\sim 30.5$ $M_\odot$), respectively (Abbott et al. 2016b). The detection of GW150914 improves our understanding of the BH-BH binaries and the BH masses 
which previously came only from the study of accreting BH systems. However, the evolution pathway toward the GW150914-like binary mergers is still under debate.
A number of recent works with different assumptions have explored the binary BH
formation and merger rate: Kinugawa et al. (2014) investigated the formation of binary BHs through
the population III stars (zero metallicity stars), and predicted high mass BHs (high chirp mass) as a typical outcome being consistent with the nature of GW150914. However, with different methods and arguments, Hartwig et al. (2016) and Dvorkin et al. (2016) claimed that GW150914 is unlikely formed from the population III stars. Further,
Marchant et al. (2016) and Mandel \& de Mink (2016) proposed the so-called chemically homogeneous evolution model for the formation of massive BHs. Mandel \& de Mink (2016) considered the chemically homogeneous evolution through binary tidal interactions. Their models may also have possible shortcomings -- for example, the chemically homogeneous evolution needs a high rotation of stars and would only occur in a binary with a very short orbital period (1.5 - 2.5 days) having a massive companion star ($\geq$ 40 $M_\odot$). We note that the chemically homogeneous evolution may also be realized through the binary mass transfer as suggested by Eldridge \& Stanway (2016). Finally, Belczynski et al. (2016) argued that a GW150914-like source is a natural outcome of the standard isolated binary evolution through the CE phase.

In sum, the formation and evolution of the progenitor system to realize GW150914 are still not fully known. Abbott et al. (2016d) summarized appropriate conditions of the isolated binary evolution scenario for the GW150914 as follows:  (1) the massive star wind cannot be strong, which means a low metallicity environment is needed  (see also Belczynski et al. 2016), (2) if the stars do not have high enough rotation and obtain small natal BH kicks, a successful evolution through the CE must be possible.

In this part, we use population II ($Z=0.001$) stars under our models 4-6 (other suitable conditions mentioned above are included in these models) to perform the BPS calculations in order to investigate a possible formation and evolution scenario of GW150914. With $Z=0.001$, we have higher BH-BH merger rates than those at the solar metallicity (see subsection 3.1). The low metallicity leads to large enhancement of the BH-BH merger rate, while the CE parameters and eccentricity also have a factor of few effect on the binary BH merger rate (section 3.3). We notice that this metallicity dependence is not clearly seen in our standard model.

In our calculations at the solar metallicity ($Z = 0.02$), the BH-BH system having the primary and secondary masses in the range observed for GW150914 is very hard to realize. Figure 8, showing the chirp mass distribution for the low metallicity model, demonstrates that a larger number of massive BH-BH binaries can be formed from Pop II stars (Fig. 8) than from Pop I stars (Fig. 3). For Models 4--6 under the low metallicity condition, a number of GW150914-like BH-BH systems are formed. Namely, a number of GW150914-like BH-BH systems can be produced from the low metallicity, Pop II stars. Our results thus support the possible origin of GW150914-like progenitors as proposed by Belczynski et al.(2016) and Abbott et al. (2016d). Below, we will investigate this issue further, by addressing how different prescriptions for $\lambda$ and $q_{\rm cr}$ affect the mass distributions of each component of the BH-BH binaries.

The nature of GW151226 is very different from that of GW150914. GW151226 has a chirp mass around $\sim 8.9$ $M_\odot$, being similar to the BH masses found in the X-ray binaries (e.g. Ozel et al. 2010). A question is if these two GW sources are evolved through the same route, and what is then is the origin of the difference. In Figure 9, we investigate mass distribution of the merging BH-BH binaries for four different models and two different metallicities.
By comparing the models with different metallicities in Figure 9, it is indeed seen that GW151226-like systems are typically expected at the solar metallicity (Pop I) while it is more likely to have GW150914-like systems at the lower metallicity (Pop II). We note that at low metallicity the BH-BH binary systems typically have a more massive `secondary' than the primary\footnote{In our definition, the primary is a more massive star at the formation of the binary system. We note that this inversion in the binary masses is also found by Eldridge \& Stanway (2016)}. This is likely caused by the less significant wind mass loss and more significant mass transfer for the lower metallicity environment. 

The delay time distribution of the BH-BH merger is shown in the left panel of Figure 10, adopting $\lambda_{\rm e}$ and $q_{\rm cr1}$. It is seen that the delay time distribution of the BH-BH mergers is indeed insensitive to the prescriptions of the binary evolution parameters such as $\lambda_{\rm e}$ and $q_{\rm qr}$. Furthermore, the shape of the delay time distribution does not evolve significantly with metallicity. As such, we expect that some fraction of the detected BH-BH systems would come from the low-metallicity, massive binary systems, once the combined effects of the high star formation rate and large chirp masses of the expected mergers at low metallicity are taken into account. This supports the idea that the GW150915-like systems represent the low-metallicity binary evolution.

The left panel of Figure 11 shows how different combinations of $\lambda$ and $q_{\rm cr}$ affect the resulting mass distributions of the BH-BH systems. We see that the treatment of $q_{\rm cr}$ does not have any clear effect. On the other hand, the binding energy parameter ($\lambda$) has non-negligible consequence on the property of the BH-BH systems. With $\lambda_{\rm g}$, we still result in massive BH-BH system at low metallicity, and the mass distribution of the primary BH component is similar to the case with $\lambda_{\rm e}$ ($\sim 6 - 40 M_\sun$). However, with $\lambda_{\rm g}$, there is still a significant fraction of the (secondary) BH below $\sim 10 M_\odot$, which is however deficient for the case with $\lambda_{\rm e}$. Since $\lambda_{\rm e}$ is more efficient in ejecting the CE, a massive binary can survive the CE more easily. The derived BH masses of GW150914 are more consistent with the results adopting $\lambda_{\rm e}$; this suggests that the treatment of the binding energy is important to understand the nature of this event.


\subsection{Implications for the rate and property of the NS-NS mergers}

With the detection of GW170817, Abbott et al.(2017c) obtained the Galactic NS-NS merger rate of $\sim 133^{+276}_{-105}$ ${\rm Myr}^{-1}$. This is much larger than the median value of 21 ${\rm Myr}^{-1}$ as previously estimated by Kim et al. (2015) 
indirectly from the observed NS-NS systems. The BH-BH merger rate obtained through statistics of the detected GW events is derived as $\sim 1 - 22$ ${\rm Myr}^{-1}$ by the LIGO GW detections O1 (Abbott et al. 2017a). The detected BH-BH mergers show diversity in their properties (especially the BH masses) as represented by GW150914 and GW151226. In this section, we discuss what implications these observational constraints have on the nature of the formation and evolution of the DCBs, by making the best use of our unified treatment of these systems in a single BPS scheme.

As we discussed, the BH-BH merger rate is relatively insensitive to the binary parameters (\S 3.1). There is enhancement of the BH-BH merger rate for $\lambda_{\rm e}$, $q_{\rm qr 1}$, and/or short initial orbital period. However, the difference in the resulting BH-BH merger rate is up to a factor of a few for different parameter sets at solar metallicity (which should dominate the number of detectable GW sources), and therefore any model can be in the observationally constrained range of the BH-BH merger rates. In other word, our BPS model provides the BH-BH rate consistent with the observationally derived rate, irrespective of the details of the binary evolution. However, the mass distributions of the BH-BH systems are affected by these binary parameters, especially by $\lambda$ (\S 3.3). If we are to explain the masses of the BHs found for GW150914 by an isolated binary evolution at low metallicity within our formalism, we require to adopt $\lambda_{\rm e}$. 

For the physically-motivated prescription adopting $\lambda=\lambda_{\rm e}$ and $q_{\rm cr} = q_{\rm cr3}$ with the tighter initial orbital period distribution, which also provides a reasonable explanation of GW150914 (e.g., see the above discussion for $\lambda_{\rm e}$), the predicted NS-NS merger rate becomes high under both solar and low metallicity (see \S 3.1). The detected BH-BH merger rate and NS-NS merger rate by LIGO (O1 and O2) can then be explained in the same context in our study (e.g., the rates under our model 4 with case 3 or case 1), with a note that the solar metallicity systems should dominate the number of the BH-BH and NS-NS merger systems in the GW detectable horizon \footnote{We note that in the original version of the manuscript which had been posted on arXiv before GW170817 was announced and the updated NS-NS rate became available, we indeed emphasized the discrepancy between our high NS-NS rate with $\lambda_{\rm e}$ (to be consistent with GW150914) and the NS-NS merger rate indirectly obtained by Kim et al. (2015).}.

Several groups have derived various NS-NS merger rates and BH-BH merger rates in the same context using their own BPS codes (Mennekens \& Vanbeveren 2014; de Mink \& Belczynski 2015; Eldridge \& Stanway 2016). As a recent example, we note that Chruslinska et al. (2017) also examined if the BH-BH rates and NS-NS rates are consistently explained within the same context. They provide three such models; models C, C+P, and NK2. The latter two models seem to predict too high local BH-BH merger rates. The BH-BH rate and NS-NS rate in their model C at the solar metallicity are covered by our models. However, we note that the predicted behavior as a function of the metallicity is very different, as their model predicts lower NS-NS merger rate for lower metallicity, opposite to our work and some of previous works (e.g., Hurley et al. 2002; Eldridge \& Stanway 2016). This highlights the difficulty in comparing results obtained by different groups by using different BPS codes under different assumptions and metallicities. In our work, we have tried to test different prescriptions within the same BPS scheme so that we can identify how specific physical process can affect the outcome. More importantly, in this paper, we investigate not only the BH-BH and NS-NS rates, but also the mass distributions of these systems so that we make the best use of the available information from the observed events.

How the high NS-NS rate obtained with the detection of GW170817 is realized in the binary evolution scenario is indeed controversial. Kruckow et al. (2018) recently derived lower local DNS merger rate (with an upper limit $\sim 39$ ${\rm Myr}^{-1}$) in their theoretical work. Giacobbo \& Mapelli (2018) showed that the results of their model can match to the latest NS-NS rate if unusually high CE efficiency parameter ($\alpha_{\rm CE} $) would be adopted. In this paper, we show that the BH-BH and NS-NS rates are consistently explained with the physically motivated prescription of $\lambda = \lambda_{\rm e}$ (with $\alpha_{\rm CE} \leq 1$), which further explains the diversity in the nature of the detected BH-BH GW systems. 

The delay time distributions of the NS-NS mergers in our BPS simulations have a peak around 1 Myr (see the right panel of Figure 10), similar to other study such as Chruslinska et al. (2017). We note that our standard model (with, e.g., constant $\lambda$) results in a longer characteristic delay time for the NS-NS merger. The short delay time in our models is mostly driven by the efficient CE ejection and survival of the NS-NS systems in close orbits. Such a short delay time distribution is also favored for production of the r-process elements in the early phase of the Milky Way (Beniamini et al. 2016). 

As we have done for the properties of the BH-BH mergers, considering not only the NS-NS merger rate but also their properties will lead to additional constraints on the astrophysical nature of massive binary evolutions. The chirp mass (1.188$M_\sun$) of GW170817 derived by LIGO is reproduced by our study. We find that the treatment of the mass transfer ($q_{\rm cr}$) can be important in the mass distributions of the NS-NS systems, while the treatment of the binding energy ($\lambda$) has negligible influence (the right panel of Figure 11). If we adopt the spin-dependent mass-transfer efficiency ($q_{\rm cr3}$), the masses of the NSs in the binary are distributed in a relatively wide range, with a substantial fraction of systems having nearly equal masses in the binary components. On the other hand, adopting $q_{\rm cr1}$, the NS-NS systems tend to have a small mass ratio, where one of the two NSs has the mass clustered around $\sim 1 M_{\odot}$. As such, the mass transfer prescription is important in determining the property of the NS-NS systems. Unfortunately, the mass ratio derived for GW170817 has too large uncertainty (Abbott et al. 2017a) to discriminate between the different mass transfer models, but we expect that this issue can be tested by an increasing number of the detected Galactic NS-NS systems and future GW detections.

\section{Conclusions and Discussion}

With the Monte Carlo BPS calculations, we have studied binary evolutionary paths toward detached double compact star binaries (BH-BH, BH-NS, NS-NS binaries). In doing this, we have investigated effects of different recipes to simulate key binary evolution processes. Our calculations highlight that treating the CE parameters in a physically motivated way by considering all possible contributions by the different forms of the energies (i.e. gravitational energy, internal energy and entropy of the envelope) is crucial to study massive star binary evolution and formation of DCBs.  Furthermore, importance of studying the physical parameter such as the critical mass ratio related to the mass transfer is shown with the different results by adopting three different mass transfer processes in this study. Besides, important roles of SN mechanism, mass loss, metallicity, initial orbital period distribution and other key issues in the massive star binary evolution are emphasized. Main conclusions of this paper are summarized as follows:

\begin{enumerate}
\item Our BPS simulations tend to result in higher NS-NS and BH-BH merger rates with the binding energy prescription $\lambda_{\rm e}$ than the other forms of $\lambda$, as this lets more binaries survive from the CE phase. This also leads to a top-heavy mass function of the BH-BH systems at low metallicity.

\item The prescription of the critical mass ratio ($q_{\rm cr}$ which determines the mass transfer) does not affect the property of the BH-BH systems, but it turns out to be an important function to determine the relative rate of the BH-BH mergers to the NS-NS mergers as well as the property of the NS-NS systems. When the spin is taken into account ($q_{\rm cr3}$), the mass distribution of the NS-NS systems becomes wide, including nearly-equal masses of the binary components. On the other hand, if the mass-transfer efficiency is set constant ($q_{\rm cr1}$) , the resulting NS-NS systems tend to have imbalance in the masses of the binary components. 

\item We find that different combinations of different parameters give different results, and our different assumptions could change the NS-NS merger rate more than a factor of 10. The combination of $\lambda_{\rm e}$ and tighter orbital period distribution is needed to produce more BH-BH (with $q_{\rm cr} = q_{\rm cr1}$) and NS-NS (with $q_{\rm cr} = q_{\rm cr3}$) systems for a given metallicity.
\end{enumerate}

Our study highlights the importance of studying the BH-BH mergers and NS-NS mergers self-consistently in the same scheme, including both the rates and properties of individual events. We apply our results to the observations and obtain the following implications:

\begin{enumerate}
\item  The origins of GW150914 and GW151226 are consistent with the isolated binary evolution model through the CE evolution, where the main difference in the nature (masses) is the metallicity at the formation of the binary systems. In this interpretation, GW150914 can be explained by a Pop II system while GW151226 is from a Pop I system.

\item In this context, the physically motivated prescription for the binding energy parameter ($\lambda_{\rm e}$) is favored to explain the large masses of both primary and secondary BHs derived for GW150914.

\item The efficient CE ejection ($\lambda_{\rm e}$) as constrained by the BH-BH merger systems naturally (and inevitably) leads to a high NS-NS merger rate as is consistent with the observational constraint from the detection of GW170817.

\item The mass ratio of the NSs in GW170817 is unfortunately not-well constrained, but consistent with our BPS model. In the future, precise determination of the mass ratio in the NS-NS systems will hopefully lead to a new constraint on the mass transfer mechanism in the massive binaries.

\end{enumerate}

\begin{acknowledgements}

This work was funded by JSPS International Postdoctoral fellowship of Japan (P17022), JSPS KAK- ENHI grant no. 17F17022. The work was also supported by JSPS KAKENHI (No. 26800100, 17H02864, 18H04585, 18H05223) from MEXT and by WPI Initiative, MEXT, Japan. The authors thank the Kyoto University Foundation for their support. We also thank the referee for his/her useful and constructive comments to improve the manuscript.
\end{acknowledgements}


\begin{center}
References
\end{center}


Abadie, J., et al. 2010b, Classical and Quantum Gravity, 27, 173001

Abbott, B. P., et al. 2016a, arXiv, arXiv:1602.03837

Abbott B. P., et al., 2016b, arXiv, arXiv:1602.03840

Abbott, B. P., et al. 2016c, arXiv, arXiv:1602.03842

Abbott B. P., et al., 2016d, arXiv, arXiv:1602.03846

Abbott B. P., et al., 2016e, arXiv, arXiv:1606.04855

Abbott B. P., et al., 2016f, arXiv, arXiv:1606.04856v2

Abbott, B. P., Abbott, R., Abbott, T. D., et al. 2017a, PRL,
118, 221101

Abbott, B. P., et al. 2017b, arXiv, arXiv:1709.09660v1

Abbott, B. P., et al. 2017c, PRL, 119, 161101, arXiv, arXiv:1710.05832

Ablimit, I., \& Li, X.-D. 2015, ApJ, 800, 98

Ablimit, I., Maeda, K. \& Li, X.-D. 2016, ApJ, 826, 53

Belczynski, K., Kalogera, V. \& Bulik, T. 2002, ApJ, 572, 407

Belczynski, K., Kalogera, V., Rasio, F. A., Taam, R. E., Zezas,
A., Bulik, T., Maccarone, T. J., \& Ivanova, N. 2008, ApJS, 174,
223

Belczynski, K., Bulik, T., Fryer, C., et al. 2010, ApJ, 714, 1217

Belczynski, K., Repetto, S., Holz, D., O'Shaughnessy, R., Bulik, T.,
Berti, E., Fryer, C., \& Dominik, M. 2015, arXiv, arXiv:1510.04615

Belczynski, K., Holz, D., Bulik, T., \& O'Shaughnessy, R.
2016, arXiv:1602.04531v1

Beniamini, P., Hotokezaka, K. \& Piran, T. 2016, ApJL, 829, L13

Cantiello, M., Yoon, S. C., Langer, N. \& Livio, M. 2007, A\&A, 465, L29

Chruslinska, M., Belczynski, K., Klencki, J. \& Benacquista, M. 2017, arXiv:1708.07885

Clausen, D., Piro, A. L. \& Ott C. D. 2015, ApJ, 799, 190

Davidson, K. \& Humphreys, R. M. 2012, ASSL, 384, D

Davis, P. J., Kolb, U., Willems, B. \& G\"{a}nsicke, B. T., 2008, MNRAS, 389, 1563

Davis, P. J., Kolb, U., \& Knigge, C. 2012, MNRAS, 419, 287

Dewi, J. D. M., \& Pols, O. R. 2003, MNRAS, 344, 629

de Mink, S. E., Pols, O. R., \& Hilditch, R. W. 2007, A\&A, 467, 1181

de Mink, S. E. \& Belczynski, K. 2015, ApJ, 814, 58

Dominik, M., Belczynski, K., Fryer, C., Holz, D. E., Berti, E., Bulik,
T., Mandel, I. \& O'Shaughnessy, R. 2012, ApJ, 759, 52

Dominik, M., Berti, E., O'Shaughnessy, R. et al.  2015, ApJ, 806, 263

Dvorkin, I., Vangioni, E., Silk, J., Uzan, J., Olive, K., 2016, ArXiv: 1604.04288v1

Eggleton, P. P. 1971, MNRAS, 151, 351

Eldridge, J. J., \& Stanway, E. R. 2016, MNRAS, 462, 3302

Eldridge, J. J. Stanway, E. R. et al. 2017, PASA, 34, 58E

Eichler, D., Livio, M., Piran, T., \& Schramm, D. N. 1989, Nature, 340, 126

Einstein, A. 1918, Sitzungsberichte der K\"{o}niglich Preubischen
Akademie der Wissenschaften (Berlin), Seite 154-167., 154

Fryer, C. L., Belczynski, K., Wiktorowicz, G. et al. 2012, ApJ, 749, 91

Giacobbo N., Mapelli M., 2018, arXiv, arXiv:1806.00001

Han, Z., Podsiadlowski, P., Eggleton, P. P., 1994, MNRAS, 270, 121

Hansen, B. M. S., \& Phinney, E. S. 1997, MNRAS, 291, 569

Harry, G. M. and LIGO Scientific Collaboration, 2010, Classical and Quantum Gravity 27, 084006

Hartwig T., Volonteri M., Bromm V., Klessen R. S., Barausse
E., Magg M., Stacy A., 2016, ArXiv: 1603.05655

Heggie D.C., 1975, MNRAS, 173, 729

Hobbs, G., Lorimer, D. R., Lyne, A. G., \& Kramer, M. 2005, MNRAS, 360, 974

Humphreys, R. M. \& Davidson, K. 1994, PASP, 106, 1025

Hurley, J. R., Pols, O. R., \& Tout, C. A. 2000, MNRAS, 315, 543

Hurley, J. R., Tout, C. A., \& Pols, O. R. 2002, MNRAS, 329, 89

Iben, Jr.,I. \& Livio, M. 1993, PASP, 105, 1373

Ivanova, N. \& Chaichenets S. 2011, ApJ, 731, L36

Ivanova, N., Justham, S., Chen, X., et al. 2013, ARA\&A, 21, 59

Kiel, P. D., \& Hurley, J. R. 2006, MNRAS, 369, 1152

Kim C., Perera B. B. P. \& McLaughlin M. A. 2015, MNRAS, 448, 928

Kinugawa, T., Inayoshi, K., Hotokezaka, K., Nakauchi, D., \& Nakamura, T. 2014, MNRAS, 442, 2963

Kochanek, C. S. 2014, ApJ, 785, 28

Kochanek C. S. 2015, MNRAS, 446, 1213

Kroupa, P., Tout, C. A., \& Gilmore, G. 1993, MNRAS, 262, 545

Kruckow M. U., Tauris T. M., Langer N., Kramer M., Izzard R. G., 2018, arXiv, arXiv:1801.05433

Kulkarni, S. R., Hut, P., \& McMillan, S. 1993, Nature, 364, 421

Lipunov, V. M., Postnov, K. A., \& Prokhorov, M. E. 1997, MNRAS, 288, 245

Madau, P. \& Dickinson, M., 2014, ARA\&A, 52, 415

Mandel, I., de Mink, S. E., 2016, MNRAS, 458, 2634

Marchant, P., Langer, N., Podsiadlowski, P., Tauris, T. M., Moriya, T. J., 2016, A\&A, 588, A50

Mennekens, N., \& Vanbeveren, D. 2014, A\&A, 564, A134

Morscher, M., Umbreit, S., Farr, W. M., \& Rasio, F. A. 2013, ApJ, 763, L15

Narayan, R., Paczynski, B., \& Piran, T. 1992, ApJ, 395, L83

Nelemans, G., Yungelson, L. R., \& Portegies Zwart, S. F. 2001, A\&A, 375, 890

Nishizawa, A., Berti, E.,  Klein, A. \&  Sesana, A. 2016, arXiv:1605.01341v1

Nomoto, K., \& Kondo, Y. 1991, ApJL, 367, L19

O'Connor, E. \& Ott, C. D. 2011, ApJ, 730, 70

O'Leary, R. M., Meiron, Y., \& Kocsis, B. 2016, arXiv:1602.02809

Ozel, F., Psaltis, D., Narayan, R. \& McClintock, J. E. 2010, ApJ, 725, 1918

Paczy\'nski, B. 1976, in Structure and Evolution of Close Binary Systems, eds. P.Eggleton,
S. Mitton, \& J. Whelan (Dordrecht: Kluwer), IAU Symp., 73, 75

Petrovic, J., Langer, N. \& van der Hucht, K. A. 2005, A\&A, 435, 1013

Phinney E. S., 1991, ApJL 380, L17

Postnov, K. A., Yungelson, L. R. 2014, LRR, 17

Sana, H., et al. 2012, Science, 337, 444

Sion, E. M., Bond, H. E., Lindler, D., et al. 2012, ApJ, 751, 66

Shao, Y., \& Li, X.-D. 2014, ApJ, 796, 37

Sengupta, A. S., LIGO Scientific Collaboration and Virgo Collaboration, 2010, Journal of
Physics Conference Series, 228, 012002.

Spera, M., Mapelli, M. and Bressan, A., 2015, MNRAS, 451, 4086

Somiya, K. 2012, Classical and Quantum Gravity, 29, 124007

Stevenson, S., Ohme, F., \& Fairhurst, S. 2015, ApJ, 810, 58

Sukhbold, T. \& Woosley, S. E. 2014, ApJ, 783, 10

Tagawa, H., Umemura, M., \& Gouda, N. 2016, arXiv:1602.08767

Tanikawa, A. 2013, MNRAS, 435, 1358

van den Heuvel, E. P. J., 1968, BAN, 19, 449V

van der Sluys, M. V., Verbunt, F., \& Pols, O. R. 2005, A\&A, 431, 647

Vink, J. S., \& de Koter, A. 2002, A\&A, 393, 543

Vink, J. S., de Koter, A., \& Lamers, H. J. G. L. M. 2001, A\&A, 369, 574

Voss, R., \& Tauris, T. M. 2003, MNRAS, 342, 1169

Wang, C., Jia, K. \& Li, X.-D. 2016, MNRAS, 457, 1015

Wang, C., Jia, K. \& Li, X.-D. 2016b, RAA, 16h, 9

Webbink, R. F. 1984, ApJ, 277, 355

Woosley, S. E. 2016, arXiv, arXiv:1603.00511v1

Wyse, R. F. G. 2009, in IAU Symposium, Vol. 258, IAU
Symposium, ed. E. E. Mamajek, D. R. Soderblom, \& R. F. G.
Wyse, 11-22

Xu, X.-J., \& Li, X.-D. 2010, ApJ, 716, 114

Yungelson, L. R., Lasota, J.-P., Nelemans, G., Dubus, G., van
den Heuvel, E. P. J., Dewi, J., \& Portegies Zwart, S. 2006,
A\&A, 454, 559

\clearpage


\begin{table}

\begin{center}
\caption{Different models and cases used in our calculation }

\begin{tabular}{lllccc}
 \hline\hline
Model & $\alpha_{\rm CE}$ & Eccentricity & $n(a)$ & $\sigma$ & Wind\\
\hline
 mod1 & 1 & 0-1 & $3-10^6 R_\odot$ & 190 Km/s & wind2\\
 mod2 & 1 & 0-1 & $3-10^6 R_\odot$  & 265 Km/s & wind2\\
 mod3 & 1 & 0-1 & $3-10^6 R_\odot$ & 265 Km/s & wind1\\
 mod4 & 1 & 0 & $3-10^4 R_\odot$ & 265 Km/s & wind2\\
 mod5 & Eq.(9) & 0 & $3-10^4 R_\odot$ & 265 Km/s & wind2\\
 mod6 &  Eq.(9)  & 0-1 & $3-10^4 R_\odot$ & 265 Km/s & wind2\\
 Standard & 1 & 0-1 & $3-10^6 R_\odot$ & 265 Km/s & wind1\\
 \hline\hline
Cases & $\lambda$ & $q_{\rm cr}$ & Cases & $\lambda$ &$q_{\rm cr}$\\
\hline
case1 & $\lambda_{\rm e}$ & $q_{\rm cr1}$ & case2 & $\lambda_{\rm e}$ & $q_{\rm cr2}$\\
case3 & $\lambda_{\rm e}$ & $q_{\rm cr3}$ & case4 & $\lambda_{\rm b}$ & $q_{\rm cr1}$\\
case5 & $\lambda_{\rm b}$ & $q_{\rm cr2}$ & case6 & $\lambda_{\rm b}$ & $q_{\rm cr3}$\\
case7 & $\lambda_{\rm g}$ & $q_{\rm cr1}$ & case8 & $\lambda_{\rm g}$ & $q_{\rm cr2}$\\
case9 & $\lambda_{\rm g}$ & $q_{\rm cr3}$ & Standard & 0.5 & Hurley et al. (2002) \\
\hline
\end{tabular}
\end{center}
\end{table}

\begin{table}

\begin{center}
\caption{Calculated merger rates under different models with different cases (in ${\rm Myr}^{-1}$)}

\begin{tabular}{l l l l l l l l l l}
 \hline\hline
Model/case & BH-BH & BH-NS & NS-NS & Model/case & BH-BH & BH-NS & NS-NS \\
\hline
 mod1/case1 & 15.8 & 2.03 & 45.1  & mod3/case1 & 12.8 & 3.0 & 69.5 \\
 mod1/case2 & 3.41 & 0.52 & 29.2 & mod3/case2 & 6.0 & 1.5 & 48.5 \\
 mod1/case3 & 6.01 & 6.02 & 74.0  & mod3/case3 & 10.5 & 8.1 & 76.2 \\
 mod1/case4 & 9.8 & 0.32 & 30.4   & mod3/case4 & 12.6 & 2.8 & 60.1 \\
 mod1/case5 & 6.42 & 0.12 & 24.9  & mod3/case5 & 6.1 & 2.0 & 45.1 \\
 mod1/case6 & 3.76 & 5.28 & 39.5  & mod3/case6 & 10.5 & 7.5 & 72 \\
 mod1/case7 & 10.1 & 0.21 & 23.2  & mod3/case7 & 12.6 & 1.7 & 50.4 \\
 mod1/case8 & 3.76 & 0.09 & 21    & mod3/case8 & 6.0 & 1.3 & 40.2 \\
 mod1/case9 & 6.45 & 5.01 & 36.1  & mod3/case9 & 10.5 & 7.2 & 62.3 \\
 mod2/case1 & 20.3 & 4.0 & 62.1   & mod4/case1 & 25.8 & 4.36 & 130 \\
 mod2/case2 & 3.3 & 0.9 & 34.8  & mod4/case2 & 8.22 & 3.58 & 84.8 \\
 mod2/case3 & 7.0 & 6.5 & 100    & mod4/case3 & 16.1 & 13.8 & 153 \\
 mod2/case4 & 12.4 & 0.7 & 42.1  & mod5/case1 & 24.4 & 2.86 & 100  \\
 mod2/case5 & 3.4 & 0.4 & 29.1   & mod5/case2 & 7.01 & 2.15 & 79.8 \\
 mod2/case6 & 3.6 & 5.2 & 58.5  & mod5/case3 & 16.2 & 10 & 122 \\
 mod2/case7 & 11.8 & 0.4 & 36   & mod6/case1 & 27.5 & 5.12 & 83.4 \\
 mod2/case8 & 1.8 & 0.1 & 22.1  & mod6/case2 & 8.01 & 4.38 & 57.8 \\
 mod2/case9 & 3.8 & 4.5 & 51.4  & mod6/case3 & 18.2 & 13.2 & 101 \\
 Standard & 15.1 & 1.21 & 40.3  & & & & &\\
\hline
\end{tabular}
\end{center}
\end{table}

\clearpage

\begin{figure}
\centering
\includegraphics[totalheight=3.5in,width=4.5in]{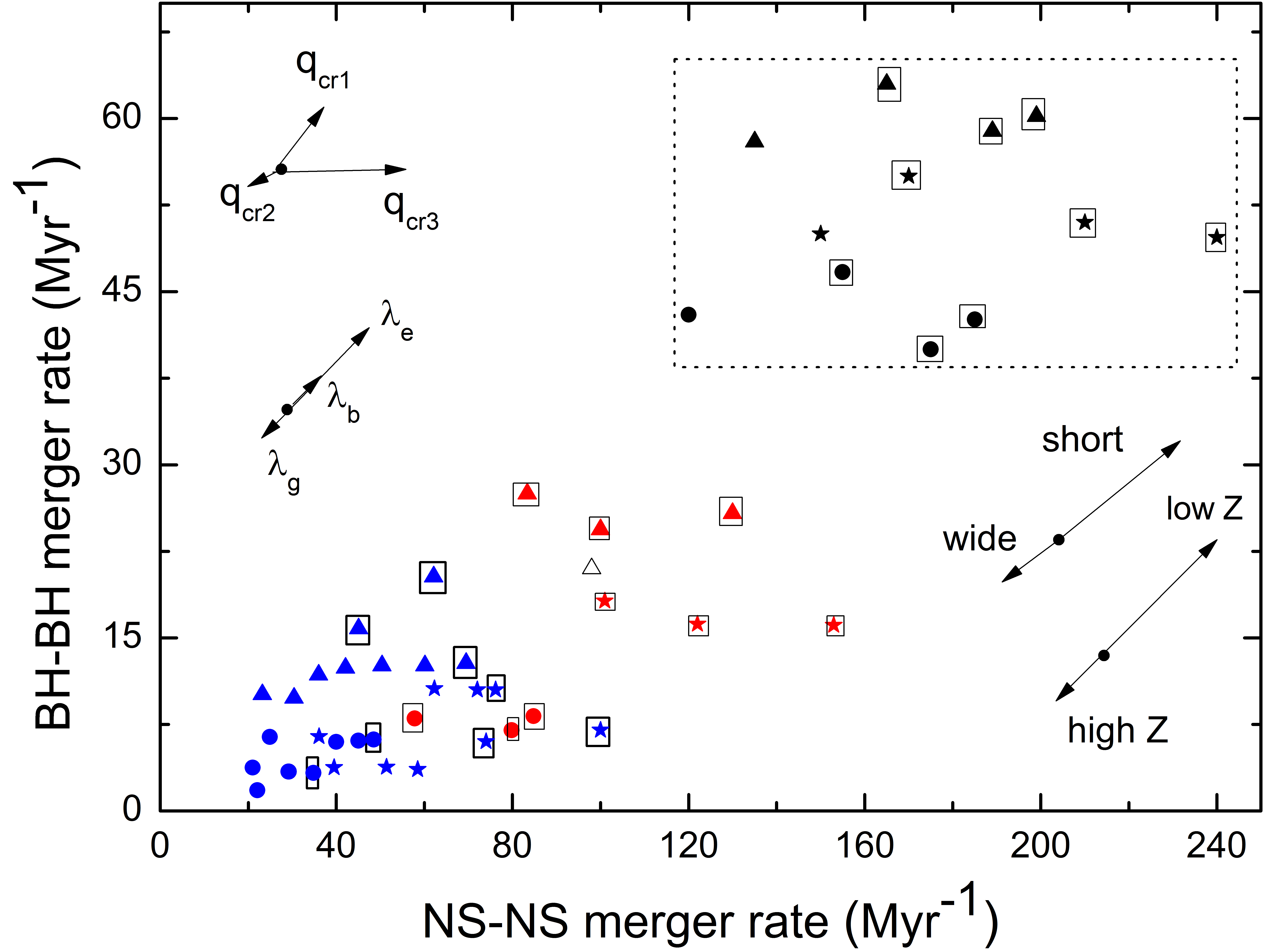}
\caption{The NS-NS merger rates versus BH-BH merger rates under different model/case and with different metallicities. Different symbols and colors are used to clarify the main factors determining the merger rates: The symbols distinguish the mass transfer prescriptions (triangles for $q_{\rm cr1}$, circles for $q_{\rm cr2}$, and stars for $q_{\rm cr3}$). Black points are for `Pop II' (Z = 0.001), which are enclosed by the dotted region. The `Pop I' models (Z = 0.02) are indicated by blue (for the wider initial orbital separation distribution) and red points  (for the tighter initial orbital separation distribution). The models with $\lambda_{\rm e}$ are shown with the box. The open triangle is for the case where LBV wind mass loss is switched off (0), and the case of LVB mass loss with $10^{-3} M_\odot{\rm{yr}}^{-1}$ is not shown as this result in nearly indistinguishable rate to the case with $1.5\times10^{-4} M_\odot{\rm{yr}}^{-1}$.}
\label{fig:1}
\end{figure}

\clearpage

\begin{figure}
\centering
\includegraphics[totalheight=4.8in,width=5.8in]{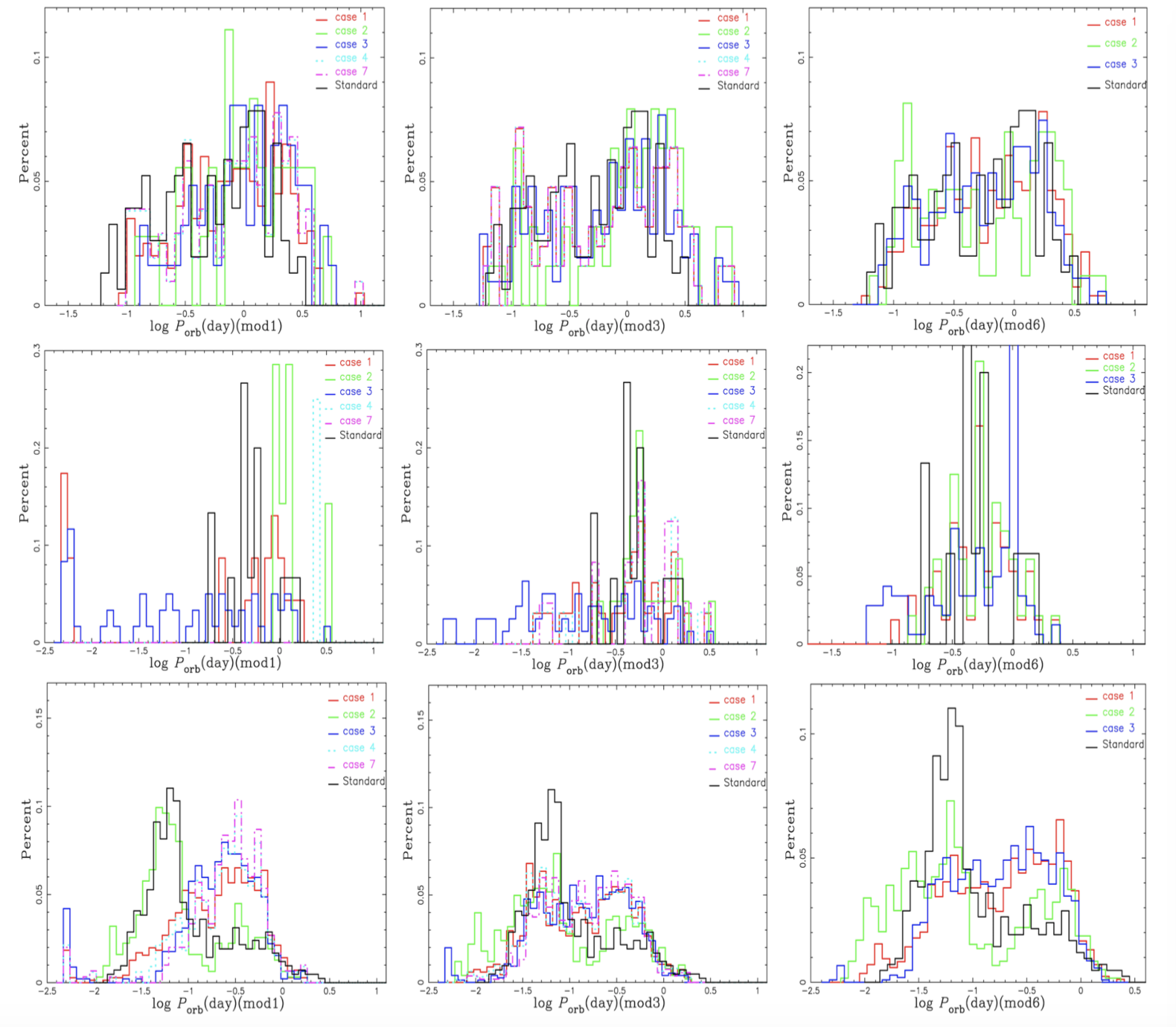}
\caption{Distribution of orbital periods of BH-BH, BH-NS and NS-NS systems (upper, middle and lower panels, respectively), for models 1, 3 and 6 (from left to right) with selected cases (see the table 1), and our standard model (the black line).}
\label{fig:1}
\end{figure}

\clearpage


\begin{figure}
\centering
\includegraphics[totalheight=4.5in,width=5.3in]{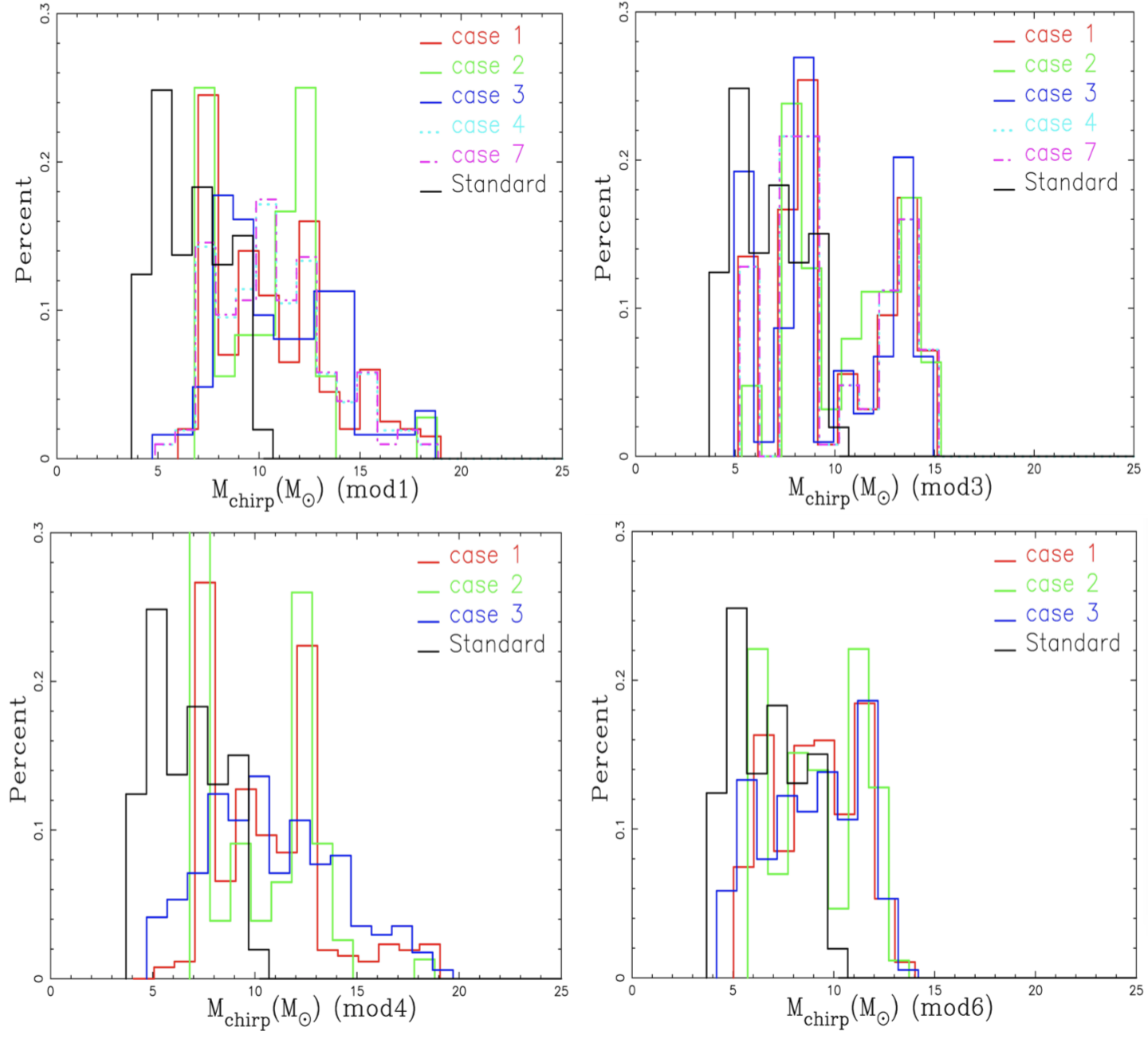}
\caption{Distribution of chirp mass of BH-BH systems, for models 1 and 3 (upper two panels) with selected cases (see table 1), models 4 and 6 (lower two panels) with 3 cases, and our standard model (the black line).}
\label{fig:1}
\end{figure}

\clearpage

\begin{figure}
\centering
\includegraphics[totalheight=4.5in,width=5.3in]{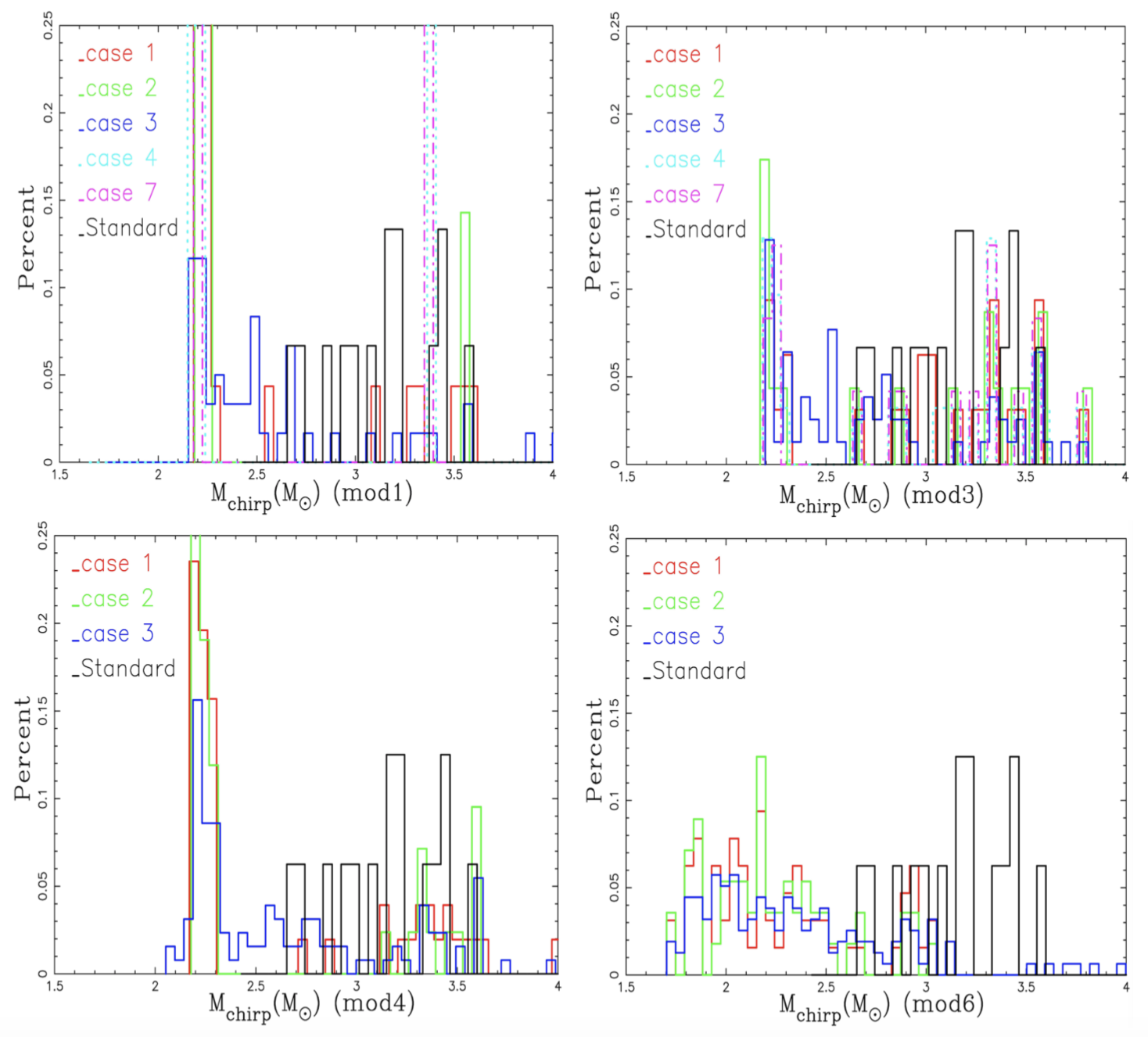}
\caption{Same as the Fig. 3, but for distribution of the chirp mass of BH-NS systems.}
\label{fig:1}
\end{figure}

\clearpage

\begin{figure}
\centering
\includegraphics[totalheight=4.5in,width=5.3in]{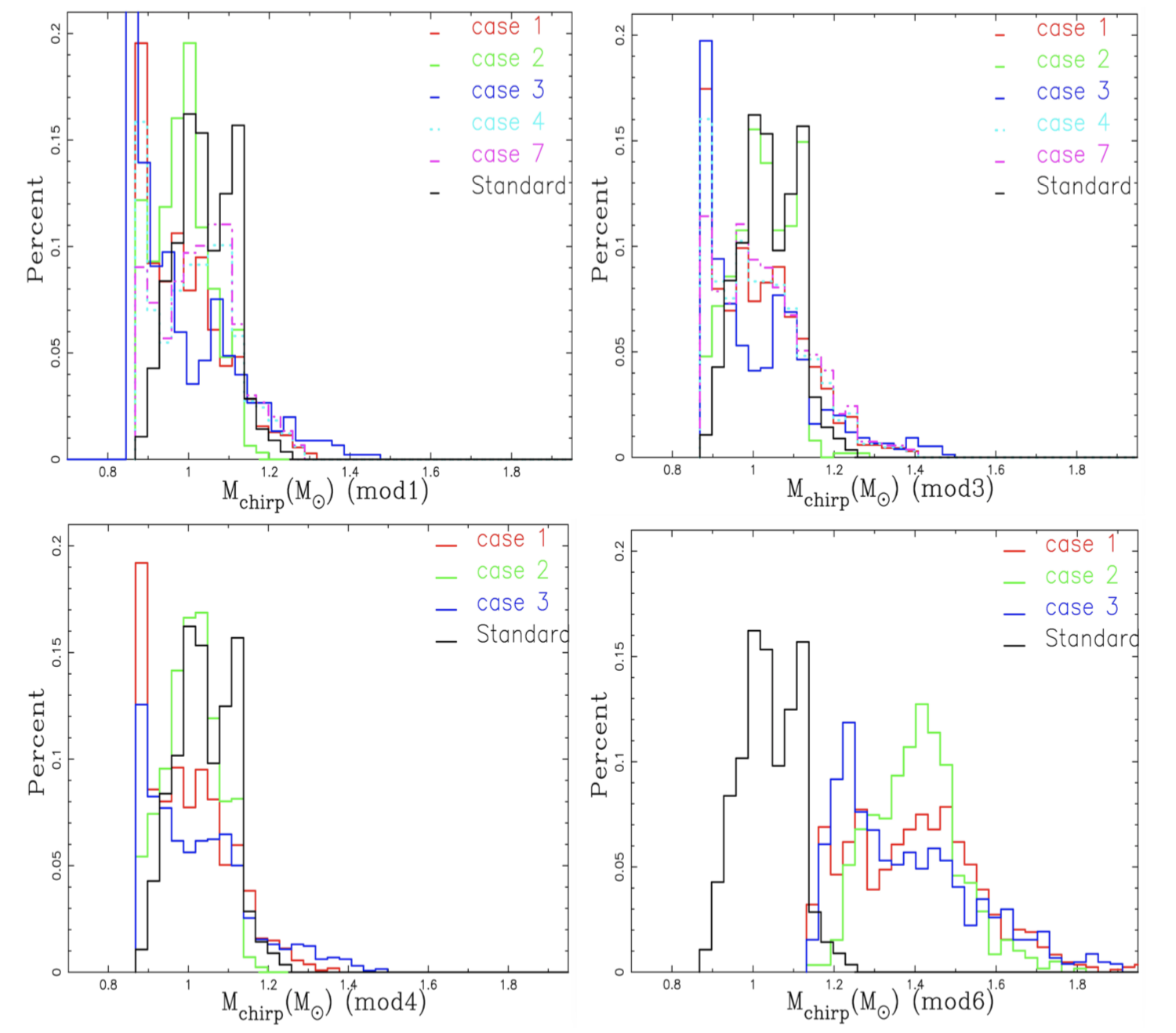}
\caption{Same as the Fig. 3, but for distribution of the chirp mass of NS-NS systems.}
\label{fig:1}
\end{figure}

\clearpage

\begin{figure}
\centering
\includegraphics[totalheight=2.4in,width=3.6in]{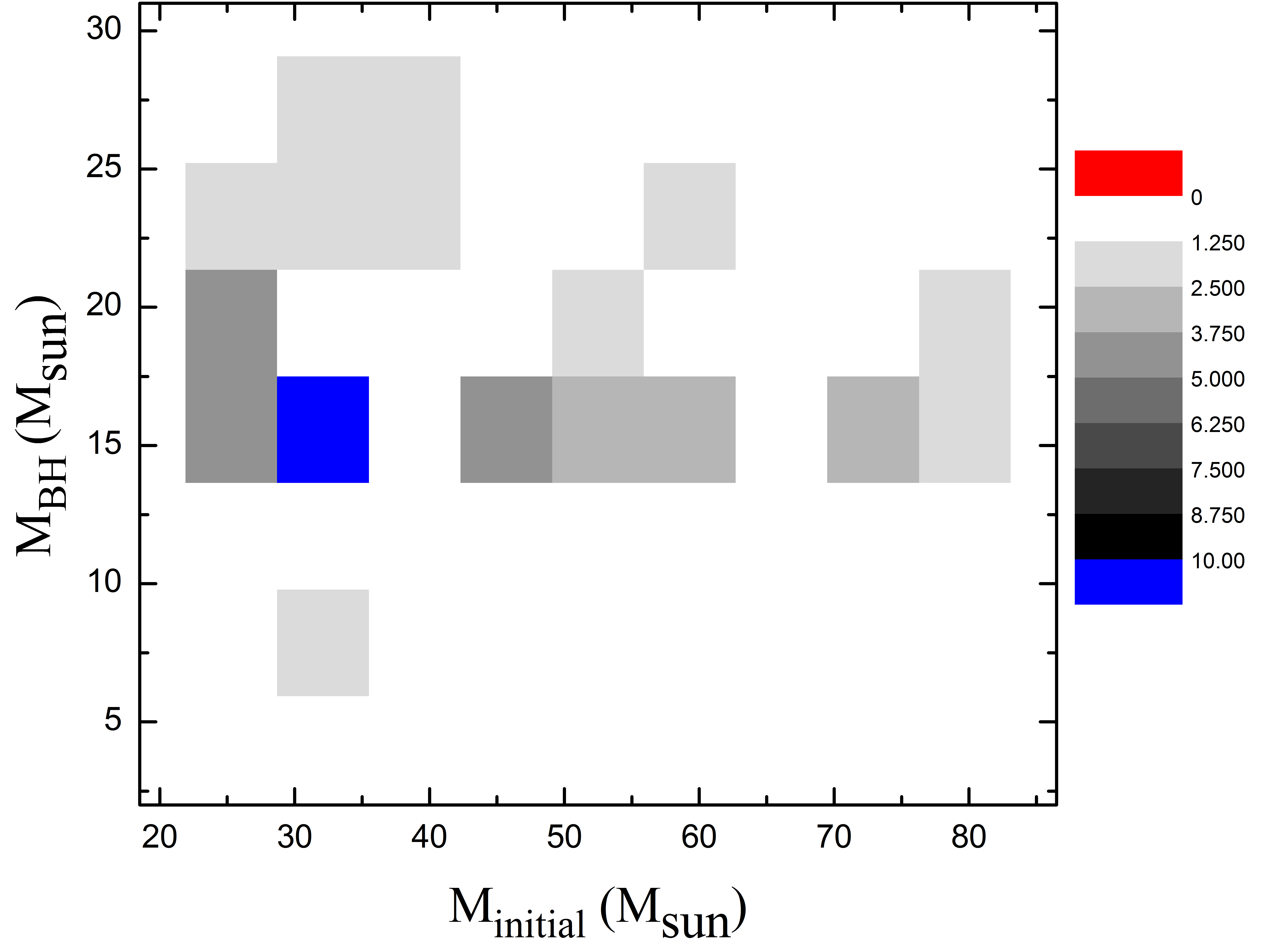}
\includegraphics[totalheight=2.4in,width=3.6in]{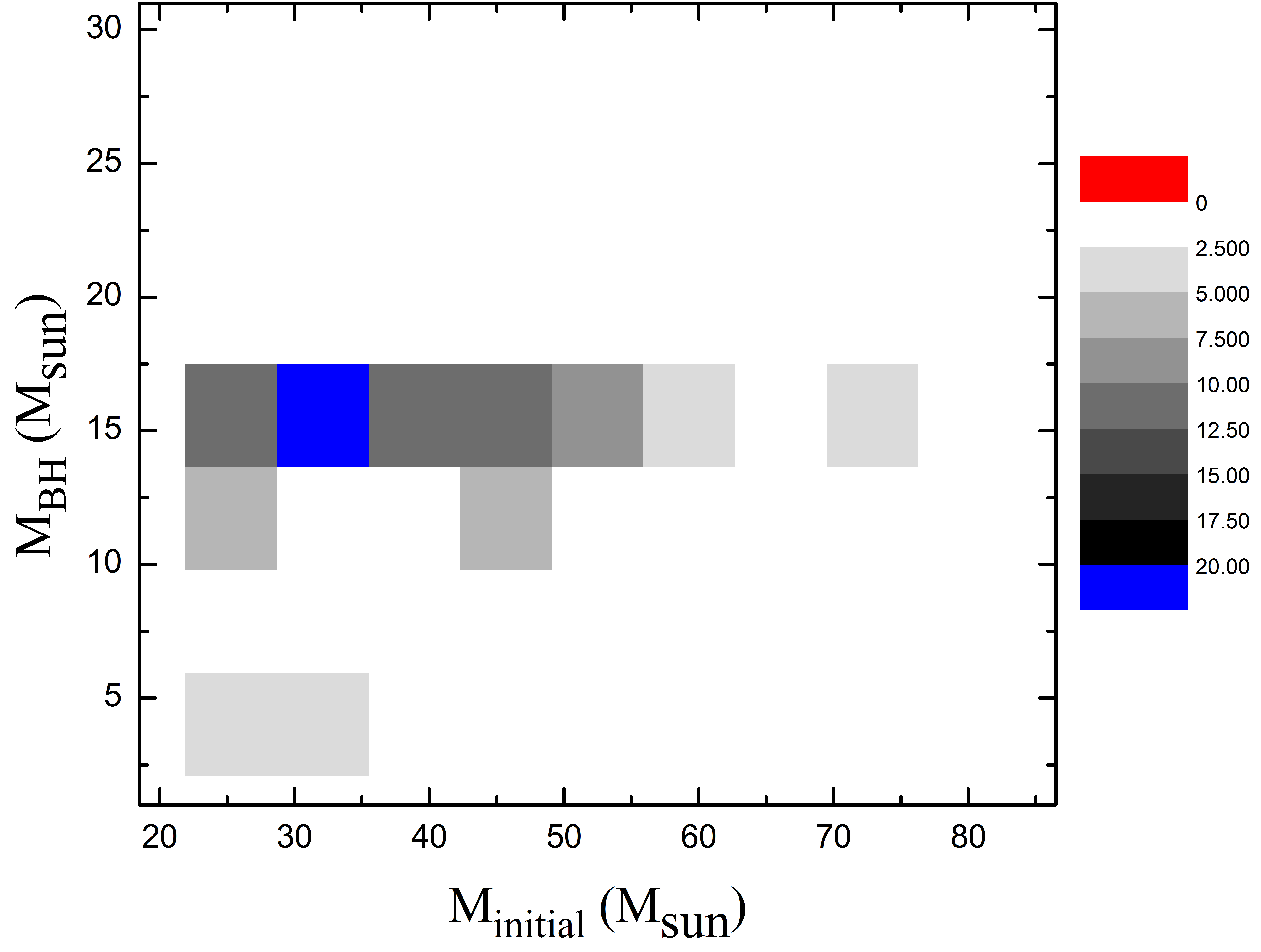}
\includegraphics[totalheight=2.4in,width=3.0in]{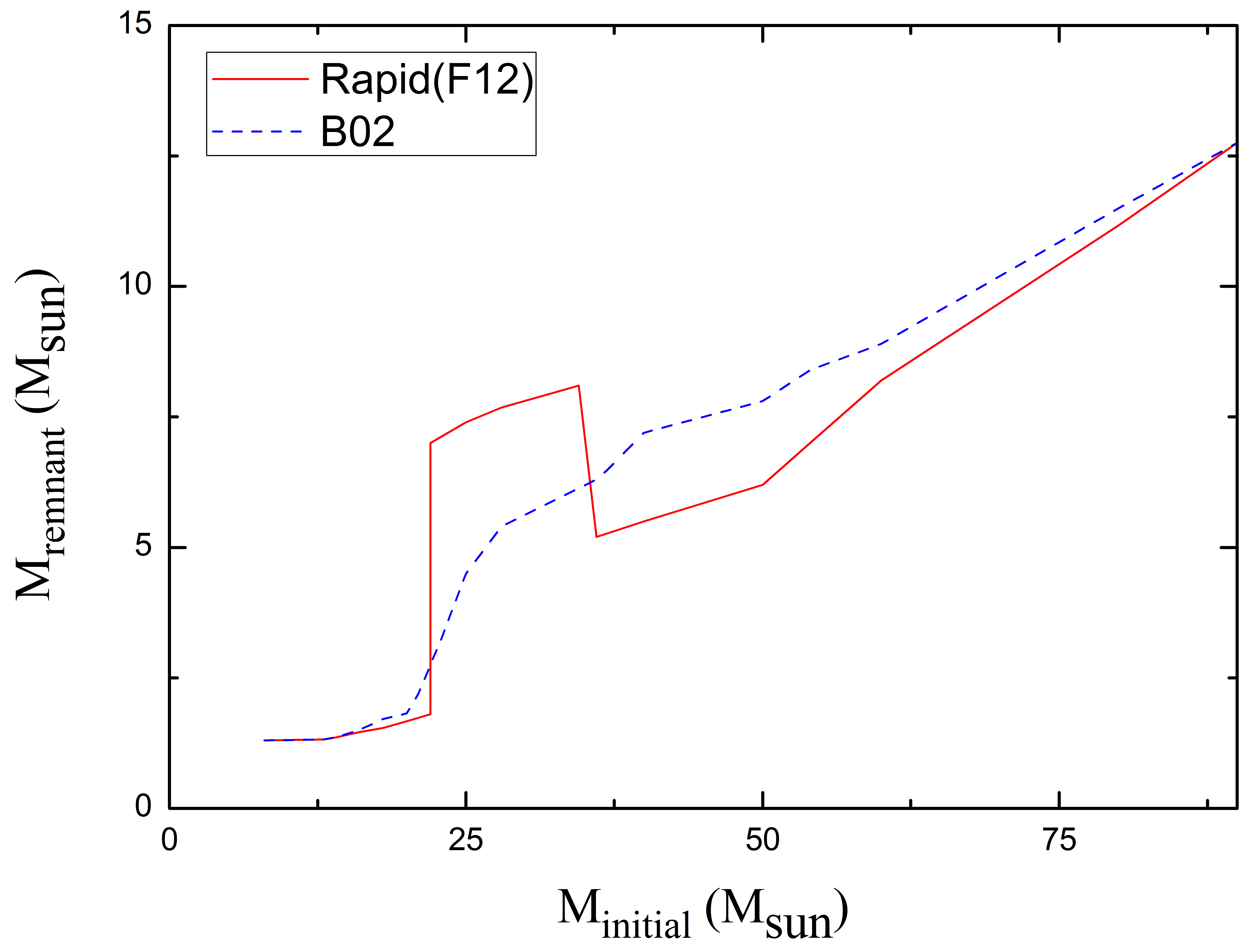}
\caption{Distribution of the primary BH mass of the BH-BH systems, for two different SN explosion models under solar metallicity and wind 2 model. The result with the SN model by Fryer et al. (2012), as generally adopted in our BPS models, is shown in the upper  panel, while the result adopting the SN model by Belczynski et al. (2002) is shown in the middle panel. The lower panel shows the remnant mass distributions from the single MS star evolution under two different SN mechanism.}
\label{fig:1}
\end{figure}

\clearpage

\begin{figure}
\centering
\includegraphics[totalheight=3.8in,width=4.2in]{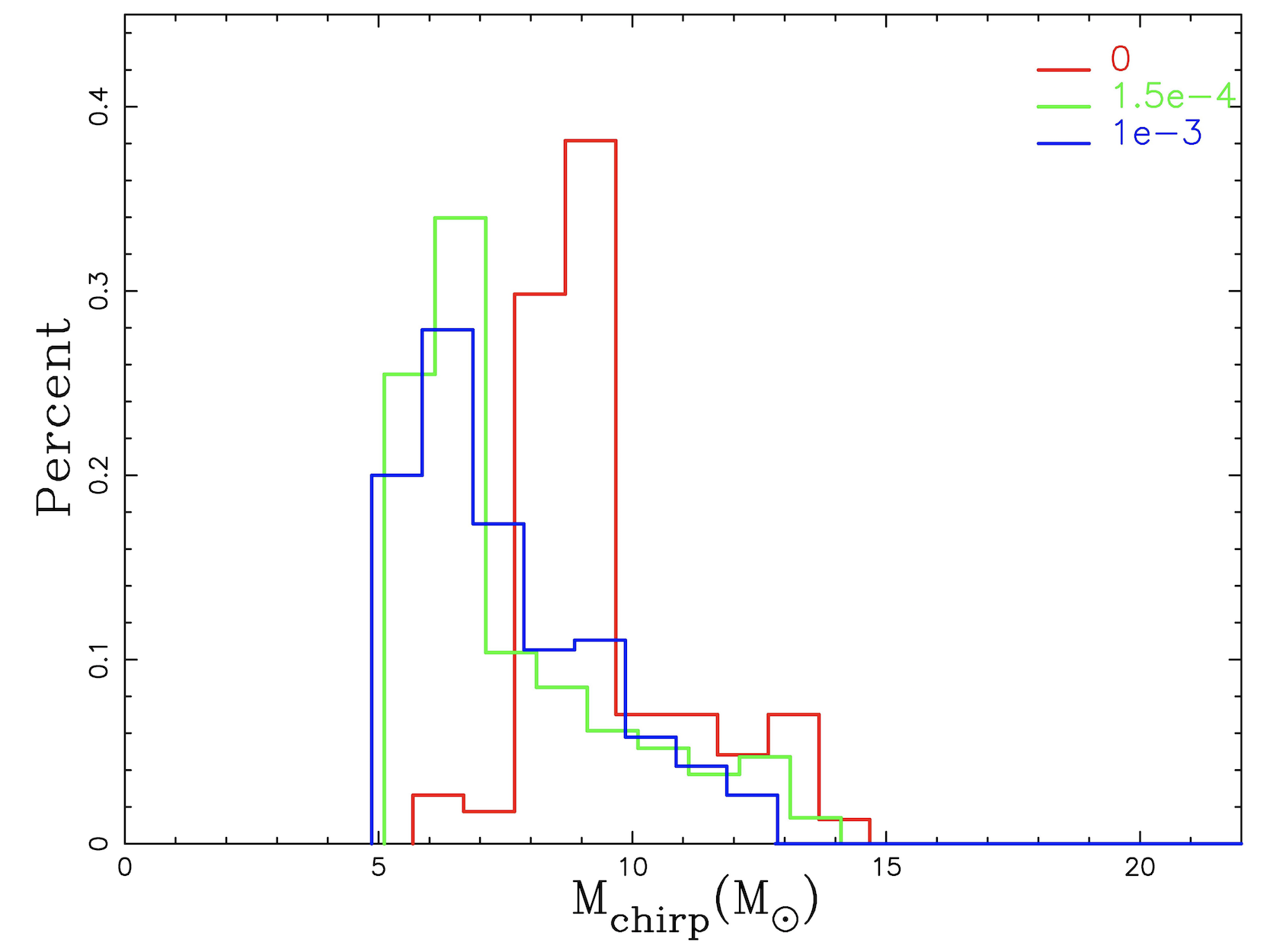}
\caption{Distribution of chirp mass under the same model (model 3 with $q_{\rm cr1}$, $\alpha_{\rm CE} = 1$, $\lambda = 0.5$ and solar metallicity) but with different wind mass loss rate for the LBVs, $\dot{M}_{\rm lbv,wind} = 0, 1.5\times10^{-4} \rm{and} 10^{-3} {\rm Myr}^{-1}$.}
\label{fig:1}
\end{figure}

\clearpage

\begin{figure}
\centering
\includegraphics[totalheight=3.0in,width=3.2in]{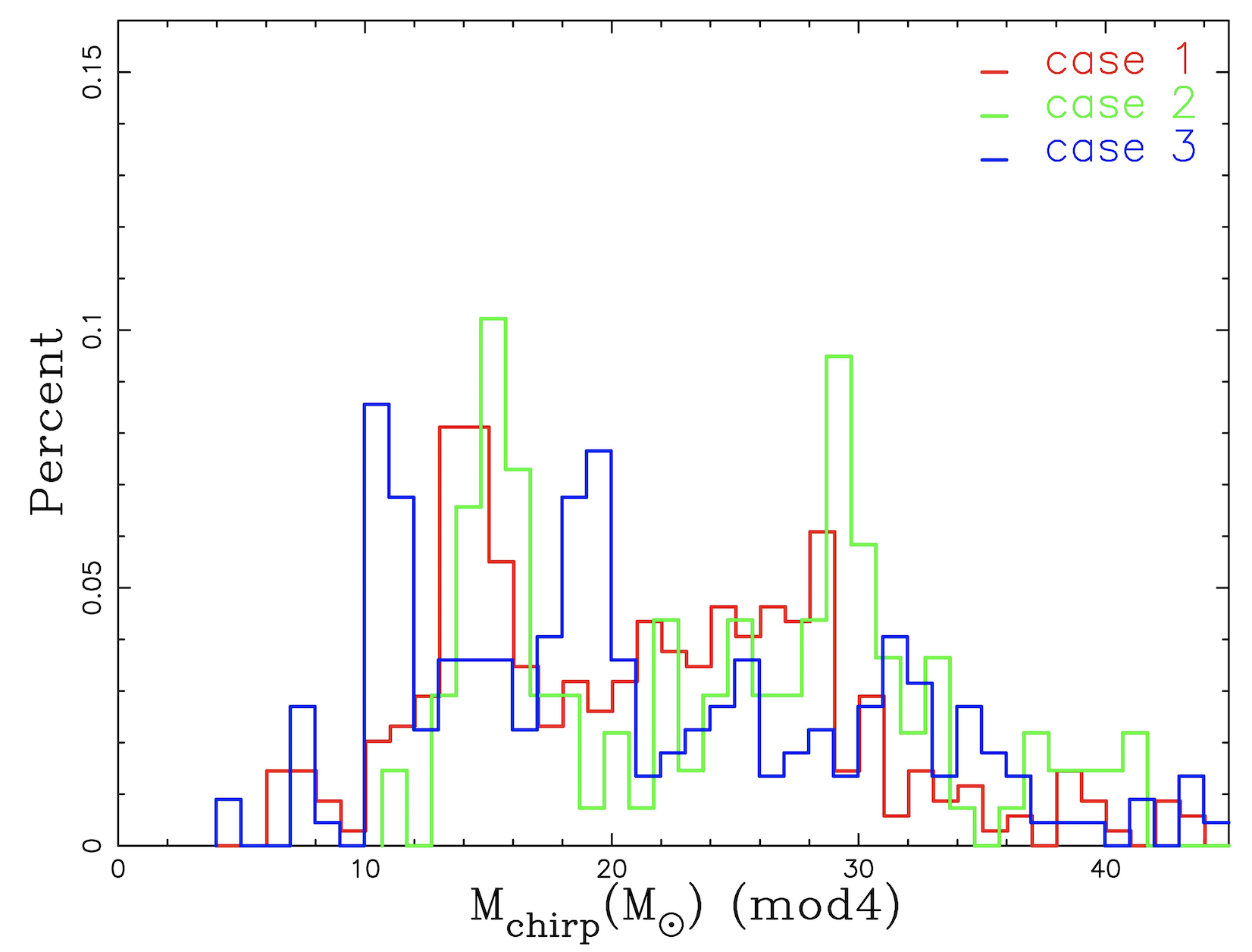}
\includegraphics[totalheight=3.0in,width=3.2in]{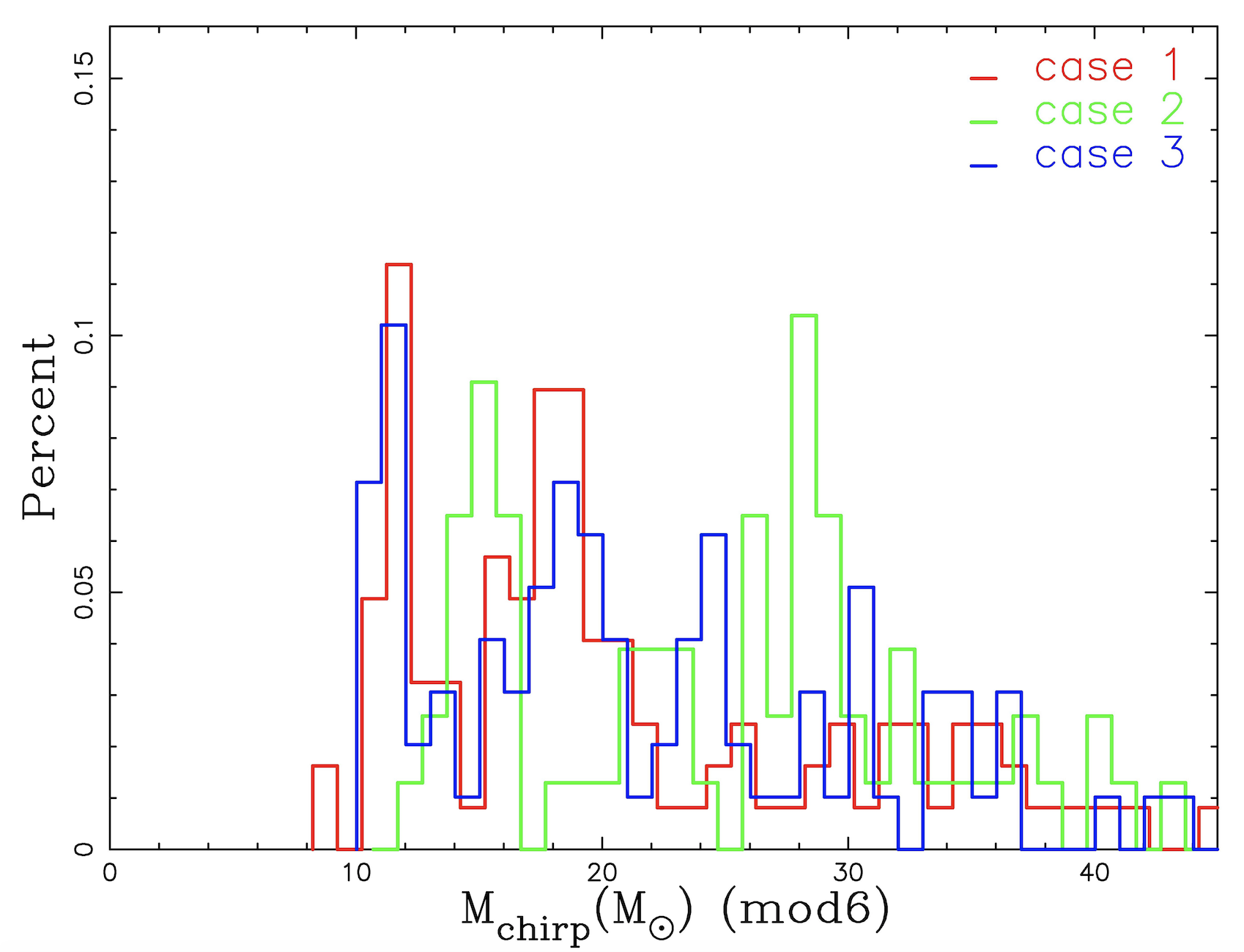}
\caption{Distribution of chirp mass under the model 4 and 6 with three cases with low metallicity (Pop II).}
\label{fig:1}
\end{figure}

\clearpage

\begin{figure}
\centering
\includegraphics[totalheight=3.0in,width=3.0in]{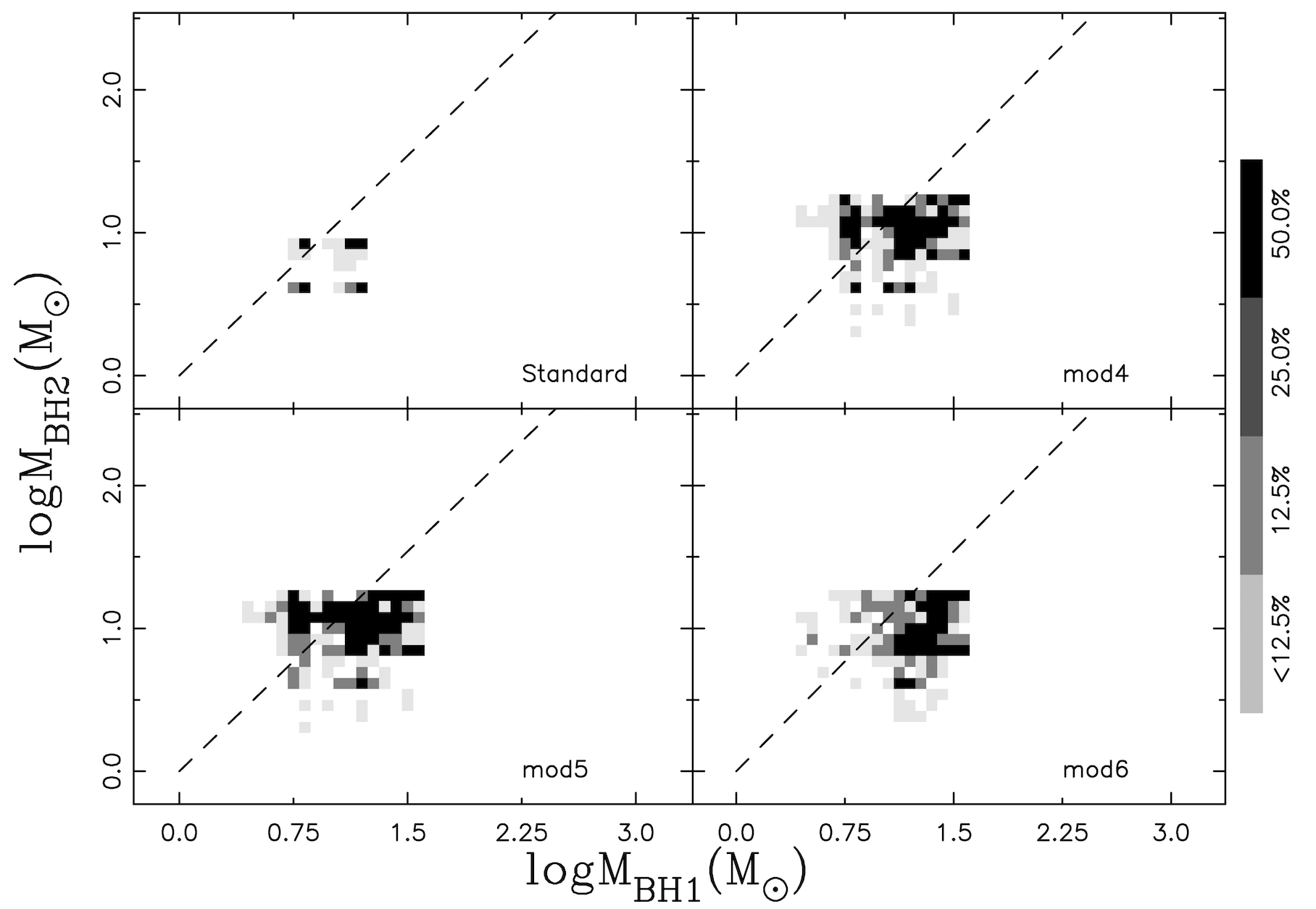}
\includegraphics[totalheight=3.0in,width=3.0in]{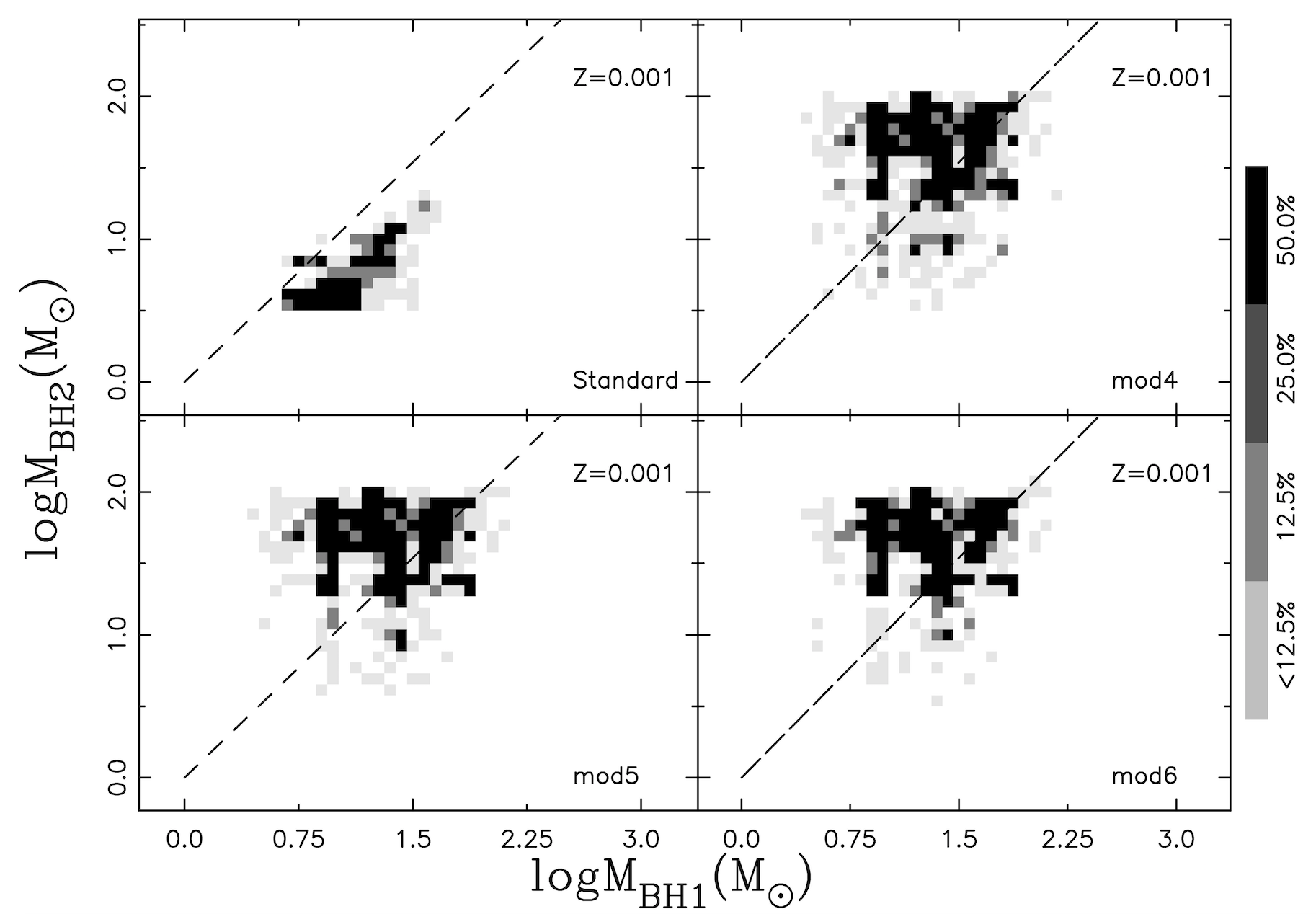}
\caption{Mass distribution of the primary and secondary BHs in the BH-BH
binaries under four models with $Z=0.02$ (left panel) and $Z=0.001$ (right panel).}
\label{fig:1}
\end{figure}

\clearpage

\begin{figure}
\centering
\includegraphics[totalheight=3.0in,width=3.0in]{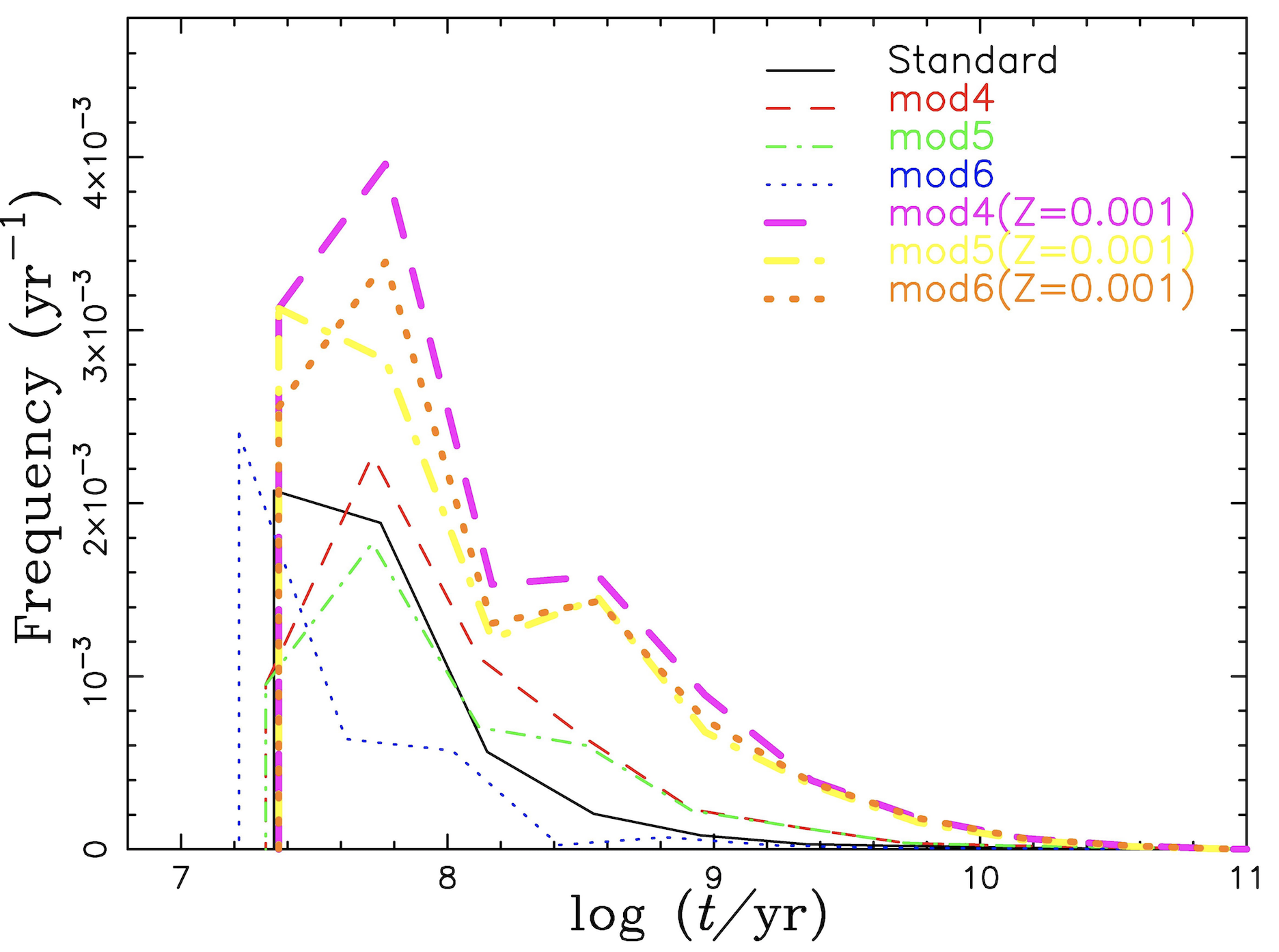}
\includegraphics[totalheight=3.0in,width=3.0in]{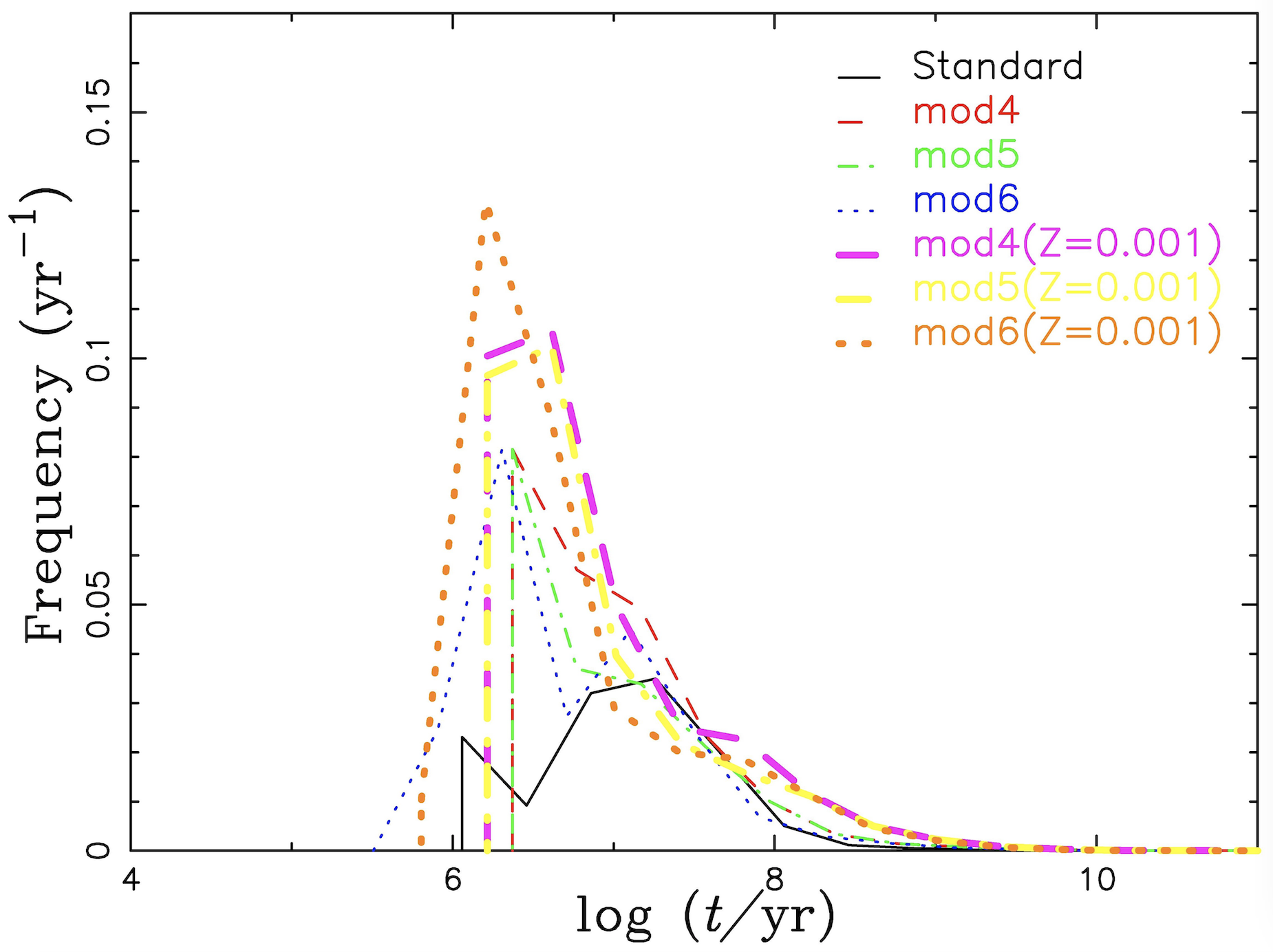}
\caption{The delay time distribution of BH-BH (left panel) and NS-NS (right panel)
mergers under standard and 4--6 models under case 1 ($q_{\rm cr1}$ and $\lambda_{\rm e}$) with $Z=0.02$ (thin lines) and $Z=0.001$ (thick lines).}
\label{fig:1}
\end{figure}

\clearpage

\begin{figure}
\centering
\includegraphics[totalheight=3.0in,width=3.1in]{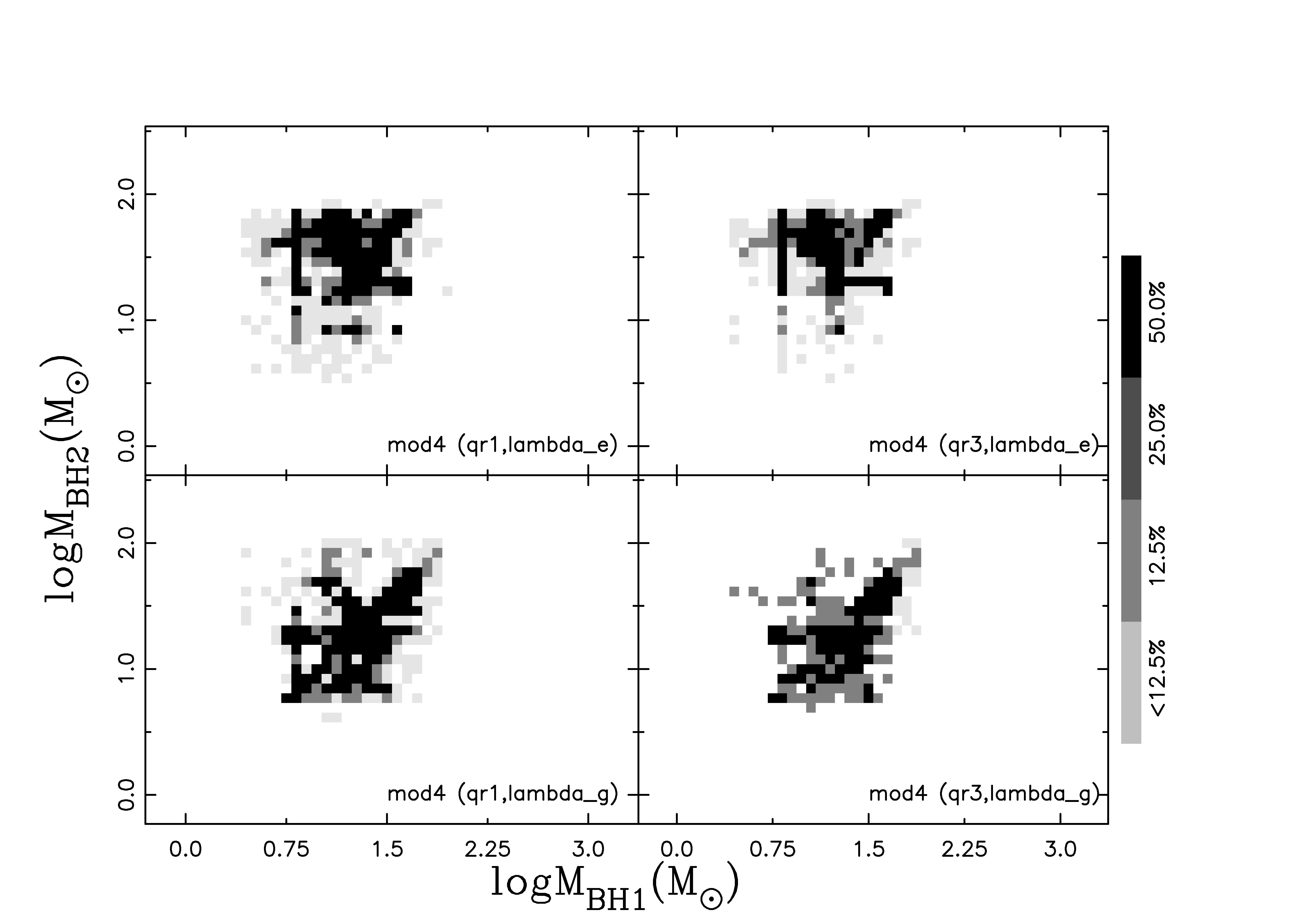}
\includegraphics[totalheight=3.0in,width=3.1in]{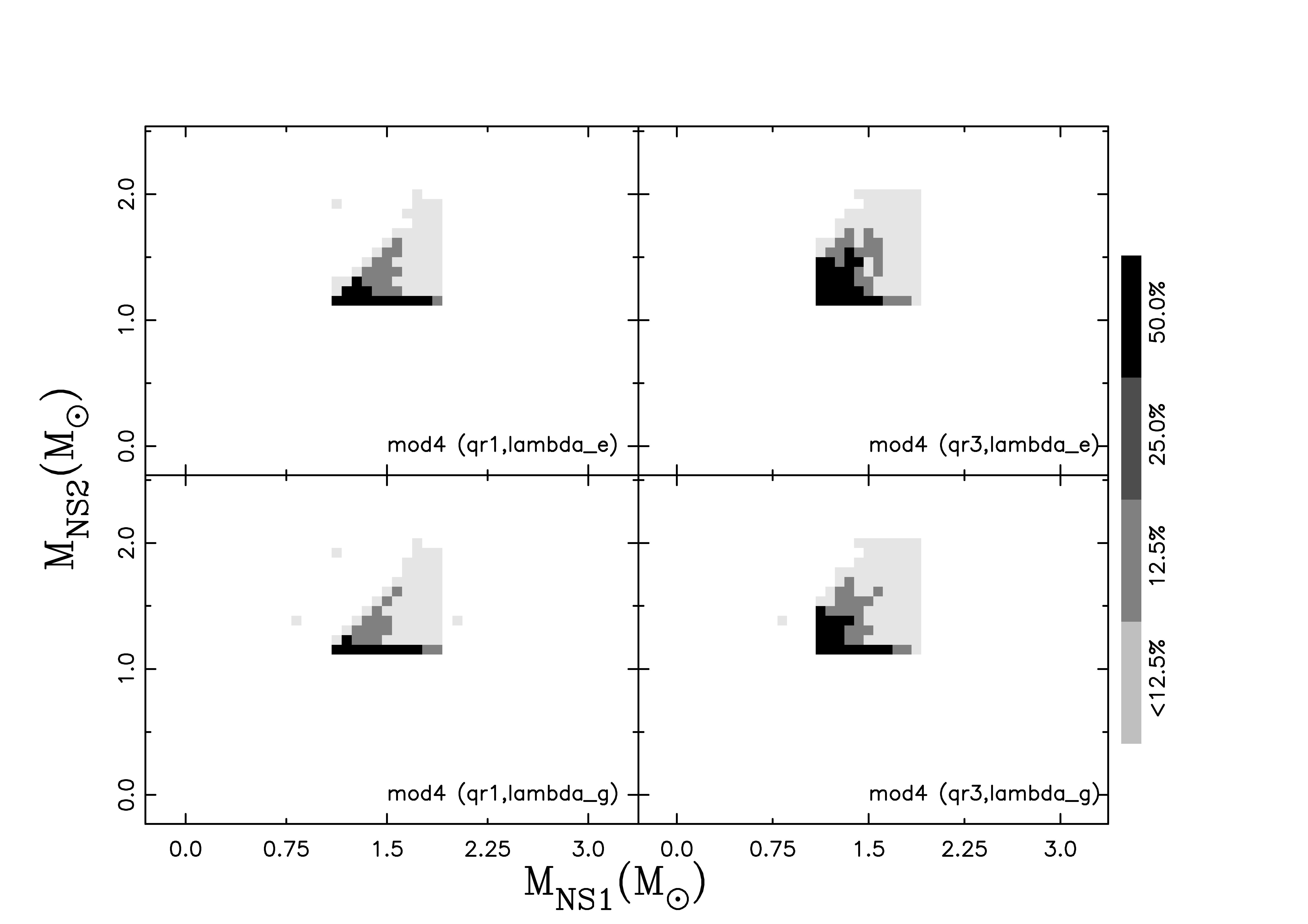}
\caption{Mass distribution of the primary and secondary in the BH-BH (left panel) and NS-NS (right panel) binaries under model 4 with different $\lambda$, $q_{\rm cr}$  and $Z=0.001$. The upper panels are for $\lambda_{\rm e}$ with $q_{\rm cr1}$ and $q_{\rm cr3}$, and the lower panels for $\lambda_{\rm g}$ with $q_{\rm cr1}$ and $q_{\rm cr3}$, respectively.}
\label{fig:1}
\end{figure}

\clearpage

\end{document}